\documentclass[%
preprint,
superscriptaddress,
preprintnumbers,
nofootinbib,
amsmath,amssymb,
aps,
prd,
]{revtex4-1}
\usepackage[utf8]{inputenc}
\usepackage{bm,amsfonts}
\usepackage{graphicx}
\usepackage{subfigure}
\usepackage{hyperref}
\hypersetup{
  colorlinks=true,
  linkcolor=red,
  citecolor=blue,
}

\begin{document}
\title{Spatially covariant gravity with a dynamic lapse function}
\author{Jiong Lin}
\email{jionglin@hust.edu.cn}

\author{Yungui Gong}
\email{Corresponding author. yggong@hust.edu.cn}
\affiliation{School of Physics, Huazhong University of Science and Technology, Wuhan, Hubei
430074, China}

\author{Yizhou Lu}
\email{louischou@hust.edu.cn}
\affiliation{School of Physics, Huazhong University of Science and Technology, Wuhan, Hubei
430074, China}

\author{Fengge Zhang}
\email{fenggezhang@hust.edu.cn}
\affiliation{School of Physics, Huazhong University of Science and Technology, Wuhan, Hubei
430074, China}

\begin{abstract}
In the framework of spatially covariant gravity, it is natural to extend a gravitational theory by putting the lapse function $N$ and the spatial metric $h_{ij}$ on an equal footing.
We find two sufficient and necessary conditions for ensuring two physical degrees of freedom (DoF) for the theory with the lapse function being dynamical by Hamiltonian analysis.
A class of quadratic actions with only two DoF is constructed.
In the case that the coupling functions depend on $N$ only, we find that the spatial curvature term cannot enter the Lagrangian and thus this theory possesses no wave solution and cannot recover general relativity (GR).
In the case that the coupling functions depend on the spatial derivatives of $N$, we perform a spatially conformal transformation on a class of quadratic actions with nondynamical lapse function to obtain a class of quadratic actions with $\dot{N}$.
We confirm this theory has two DoF by checking the two sufficient and necessary conditions.
Besides, we find that a class of quadratic actions with two DoF can be transformed from GR by disformal transformation.
\end{abstract}

\preprint{2011.05739}

\maketitle

\section{Introduction}
To explain the early and late-time accelerated expansion of the Universe \cite{Guth:1980zm,Starobinsky:1980te,Linde:1981mu,Starobinsky:1982ee,Riess:1998cb,Perlmutter:1998np}
, many extended theories of gravity have been proposed.
A natural and simple extension to general relativity (GR) is the inclusion of an extra scalar physical degree of freedom (DoF) \cite{Clifton:2011jh}.
Scalar-tensor theories, especially Horndeski theory \cite{Horndeski:1974wa} have played an important role in building models of inflation and dark energy
\cite{ArmendarizPicon:1999rj,Garriga:1999vw,
Kobayashi:2010cm,Crisostomi:2017pjs,Crisostomi:2018bsp,Frusciante:2018tvu,Lin:2020goi,Yi:2020kmq}.
Horndeski theory which contains Brans-Dicke theory
\cite{Brans:1961sx,Dicke:1961gz}, is the most general scalar-tensor theory involving up to second-order derivatives in the Lagrangian while retaining second-order field equations in four dimensions.
Horndeski theory is equivalent to the generalized Galileons which is the covariantization of a scalar theory in flat
spacetime with second-order field equations \cite{Deffayet:2011gz,Kobayashi:2011nu}.
The generalized \cite{Deffayet:2011gz} and covariant \cite{Deffayet:2009wt} Galileons break the Galileon shift symmetry in the  Galileon model \cite{Nicolis:2008in}, a covariant  model retaining the Galileon shift symmetry was constructed by localizing both Poincar$\acute{\mathrm{e}}$ symmetry and Galileon shift symmetry in Ref. \cite{Banerjee:2017qdl}.

By performing  disformal transformation \cite{Bekenstein:1992pj}
\begin{equation}
 \tilde{g}_{\mu\nu}=C(\phi,X)g_{\mu\nu}+D(\phi,X)\phi_{\mu}\phi_{\nu}
\end{equation}
on Horndeski theory, a class of degenerate higher-order scalar-tensor theories (DHOSTs) \cite{Zumalacarregui:2013pma,Bettoni:2013diz,Langlois:2015cwa,Arroja:2015wpa,Langlois:2015skt,
Crisostomi:2016czh,Achour:2016rkg,Crisostomi:2016tcp,BenAchour:2016fzp,
Langlois:2017mxy,Takahashi:2017pje} can be obtained.
DHOSTs possess higher-order derivatives in the Lagrangian while the degeneracy of the kinetic matrix indicates the existence of hidden constraints \cite{Motohashi:2016ftl,Klein:2016aiq} and consequently the possible absence of Ostrogradsky ghosts \cite{Ostrogradsky:1850fid,Woodard:2015zca}.

A timelike scalar field offers a natural spacetime foliation.
After fixing the unitary gauge, i.e. $\phi=t$, scalar-tensor theories can be regarded as spatially covariant gravity (SCG) theories with spatial diffeomorphism \cite{Khoury:2011ay,Fujita:2015ymn,Gao:2014soa,Gao:2014fra,
Gao:2018znj,Gao:2018izs,Gao:2019lpz,Gao:2019twq,Gao:2020juc,Gao:2020juc,
Gao:2020yzr,Gao:2020qxy}.
For the discussions on the construction and application of spatially diffeomorphic models, please see Refs.  \cite{Son:2005rv,Banerjee:2014pya,Banerjee:2015rca}.
Although the scalar field vanishes in the unitary gauge, the scalar-type physical DoF remains due to the breaking of spacetime diffeomorphism.
Thus to some extent, introducing the extra physical DoF beyond GR is equivalent to constructing a class of SCG.
A general Lagrangian of SCG, which depends on the lapse function $N$, spatial metric $h_{ij}$, extrinsic curvature $K_{ij}\equiv \mathsterling_{\vec n}h_{ij}/2$ and their spatial derivative $D_i$,
has been investigated in Refs. \cite{Gao:2014soa,Gao:2014fra}.
This kind of SCG includes the Horndeski theory, beyond-Horndeski theory, effective field theory (EFT) of inflation \cite{ArkaniHamed:2003uy,Creminelli:2006xe,Cheung:2007st} and the Ho$\check{\rm r}$ava gravity \cite{Horava:2009uw,Horava:2009if,Visser:2009fg,Devecioglu:2020dny}, and there are three propagating DoF.
A more general class of SCG has been studied by adding the important building block $\mathsterling_{\vec n}N$ to the Lagrangian \cite{Gao:2018znj,Gao:2019lpz}.
In GR, $N$ acts as an auxiliary field to ensure the spacetime diffeomorphism, while in SCG, there is no need to keep $N$ as an auxiliary field
and it is natural to extend a  gravitational theory by putting the lapse function and the spatial metric on an equal footing.
On the other hand, from the viewpoint of EFT,
the operator $\mathsterling_{\vec n}N$ should be included in the effective Lagrangian.
Besides, under disformal transformation or mimetic transformation \cite{Chamseddine:2013kea,Sebastiani:2016ras}, the transformed theories  may acquire $\dot{N}$.
Generally, if the lapse function is dynamic, an extra scalar mode arises and the theory has 4 DoF.
In Ref. \cite{Gao:2018znj} the authors obtained two sufficient and necessary conditions to get rid of one scalar mode through detailed Hamiltonian analysis.
An alternative derivation of the two conditions at the Lagrangian level has been studied by performing a perturbation analysis in Ref. \cite{Gao:2019lpz}.

Recently, a more aggressive attempt to seek a class of modified theories of gravity with two tensorial DoF has been made within the framework of SCG with a non-dynamic lapse function \cite{Gao:2019twq,Yao:2020tur}.
This kind of SCG naturally contains cuscuton theory \cite{Afshordi:2006ad,Afshordi:2007yx,Gomes:2017tzd,Iyonaga:2018vnu} where the scalar field is non-dynamic and the theory only propagates two tensorial DoF in the unitary gauge.
However, a further question arises: what if the lapse function is treated on an equal footing with the spatial metric?
On the other hand, motivated by field transformations, one may wonder whether there is a relation between GR and SCG with two DoF.
In this paper, as a first step, by performing detailed Hamiltonian analysis, we derive two sufficient and necessary conditions for the SCG theory with a dynamic lapse function to propagate two DoF.
The correspondence between GR and SCG with two DoF is also discussed.

The paper is organized as follows.
In Sec. \ref{sec2}, we derive the two sufficient and necessary conditions for ensuring two DoF for the theory.
Then a class of quadratic actions is constructed and the correspondence between SCG and GR is discussed in Sec. \ref{sec3}.
We conclude our paper in Sec. \ref{sec4}.

\section{Spatially covariant gravity with two physical degrees of freedom}\label{sec2}

We start with the Arnowitt-Deser-Misner 3+1 decomposition of
the four-dimensional spacetime \cite{Arnowitt:1959ah}
 \begin{equation}
   ds^2=-Ndt^2+h_{ij}(dx^i+N^idt)(dx^j+N^jdt),
 \end{equation}
where $N$ is the lapse function, $N^i$ is the shift function and $h_{ij}$ is the induced metric on the spatial hypersurfaces.
In general, the Lagrangian involves both the kinetic term of the induced spatial metric, i.e.
the extrinsic curvature $K_{ij}=\mathsterling_{\vec n}h_{ij}/2$,
and the kinetic term of the lapse function $F=\mathsterling_{\vec n}N$,
where $\mathsterling_{\vec n}$ is the Lie derivative with respect to the timelike vector $\vec n$ normal to the spatial hypersurface.
Note that to keep the spatial diffeomorphism, $N^i$ can only enter the Lagrangian through $K_{ij}$ and $F$.
In addition, terms involving the spatial derivative such as $D_m F, D_m K_{ij}$ are also allowed.
The most general action of SCG is
\begin{equation}\label{action1}
  S=\int dtd^3x \sqrt{h}N\mathcal{L}(N,h_{ij},K_{ij},F,D_i).
\end{equation}
In this section, we will derive sufficient and necessary conditions for ensuring two DoF for the theory by Hamiltonian analysis.
The Hamiltonian analysis can also be used to construct a one-to-one correspondence between the gauge and reparametrization parameters \cite{Mukherjee:2007yi}.
For the sake of Hamiltonian analysis, we introduce two auxiliary fields $A$ and $B_{ij}$ to rewrite the action \eqref{action1} as
\begin{equation}\label{action2}
  S=S_B
  +\int d^4x \left[\frac{\delta S_B}{\delta A}(F-A)+\frac{\delta S_B}{\delta B_{ij}}(K_{ij}-B_{ij})\right],
\end{equation}
where $S_B$ is obtained from the action \eqref{action1} by replacing
$K_{ij}$ with  $B_{ij}$ and $F$ with $A$.
In the case that $S_B$ doesn't involve the spatial derivative of $A$ and $B_{ij}$,
varying the action  with respect to $A$ and $B_{ij}$, we get
\begin{gather}
   \frac{\delta^2 S_B}{\delta A^2}(F-A)=0,\\
  \frac{\delta^2 S_B}{\delta B_{ij} \delta B_{kl}}(K_{kl}-B_{kl})=0.
\end{gather}
To ensure $F=A$ and $K_{ij}=B_{ij}$, we require
\begin{equation}
  \frac{\delta^2 S_B}{\delta A^2}\neq0,\
  \frac{\delta^2 S_B}{\delta B_{ij} \delta B_{kl}}\neq0.
\end{equation}
In the SCG, the 17 canonical variables are
\begin{equation}
\{\Phi_I\}=\{N^i,A,B_{ij},N,h_{ij}\},
\end{equation}
and the corresponding conjugate momenta are
\begin{equation}
  \{\Pi^I\}=\{\pi_i,p,p^{ij},\pi,\pi^{ij}\}
\end{equation}
with
\begin{equation}
\label{prconstr1}
\begin{split}
  \pi_i=0,\ p=0,\ p^{ij}=0,\ \pi=\frac{1}{N}\frac{\delta S_B}{\delta A},\\
  \pi^{ij}=\frac{1}{2N}\frac{\delta S_B}{\delta B_{ij}}.
  \end{split}
\end{equation}
From Eq. \eqref{prconstr1}, we get  the primary constraints
\begin{equation}
\{\varphi^I\}=\{\pi_i,p,p^{ij},\tilde{\pi},\tilde{\pi}^{ij}\}\approx0,
\end{equation}
where
\begin{equation}
\begin{split}
 &\tilde{\pi}=\pi-\frac{1}{N}\frac{\delta S_B}{\delta A},\
  \tilde{\pi}^{ij}=\pi^{ij}-\frac{1}{2N}\frac{\delta S_B}{\delta B_{ij}},
  \end{split}
\end{equation}
and ``$\approx$"  denotes weak equality which is valid on the subspace of phase space determined by the primary constraints.
The corresponding Lagrange multipliers are
\begin{equation}
  \{\lambda_I\}=\{\lambda^i,\Lambda,\Lambda_{ij},\lambda,\lambda_{ij}\}.
\end{equation}
The canonical Hamiltonian is obtained from the Lagrangian by a Legendre transformation
\begin{equation}
\begin{split}
  H_C=&\int d^3x (\Pi^I\dot{\Phi}_I-N\sqrt{h}\mathcal{L}_B)\\
  =&\int d^3x (NC+\pi\mathsterling_{\vec N}N+\pi^{ij}\mathsterling_{\vec N}h_{ij}),
\end{split}
\end{equation}
where
\begin{equation}
  C= \pi A+ 2\pi^{ij}B_{ij}-\sqrt{h}\mathcal{L}_B,
\end{equation}
and $\mathcal{L}_B$ is obtained from $\mathcal{L}$ in Eq. \eqref{action1} by replacing
$K_{ij}$ with  $B_{ij}$ and $F$ with $A$.
Using the constraints $p^{ij}\approx0,\ p\approx0,\ \pi_i\approx0$, integrating by parts and choosing Minkowski spacetime as boundary condition at infinity, the canonical Hamiltonian can be recast in
the form
\begin{equation}
\label{hceq2}
  H_C\approx \int d^3x (NC+N^iC_i),
\end{equation}
where
\begin{equation}
\begin{split}
  C_i=&\pi D_iN-2\sqrt{h}D_j\left(\frac{\pi^j_i}{\sqrt{h}}\right)
  +pD_iA
  +p^{kl}D_iB_{kl}\\
  &-2\sqrt{h}D_l\left(\frac{p^{lk}}{\sqrt{h}}B_{ik}\right)
  +\pi_kD_iN^k +\sqrt{h}D_k\left(\frac{\pi_iN^k}{\sqrt{h}}\right).
  \end{split}
\end{equation}
Defining the Poisson bracket
\begin{equation}
\label{pbrack1}
  [F,G]=\int d^3z \left(\frac{\delta F}{\delta \Phi_I}\frac{\delta G}{\delta \Pi^I}-\frac{\delta F}{\delta \Pi^I}\frac{\delta G}{\delta \Phi_I}\right),
\end{equation}
and using the Hamiltonian \eqref{hceq2} and the Poisson bracket \eqref{pbrack1},
we get the consistency conditions
\begin{equation}\label{consistencycondition}
  \dot{\varphi}^I(\vec x)=
  \int d^3y D^{IJ}\lambda_J+[\varphi^I(\vec x),H_C]\approx0,
\end{equation}
where the values of the Dirac matrix $D^{IJ}=[\varphi^I(\vec x),\varphi^J(\vec y)]$ are presented in Appendix \ref{app.A}.
The consistency condition of the constraint $\pi_i\approx0$ leads to three secondary constraints
\begin{equation}
 \dot{\pi}_i(\vec x)=-C_i(\vec x)\approx 0.
\end{equation}
Note that it was proved in Ref. \cite{Gao:2018znj} that the Poisson brackets between $C_i\approx0$ and
any tensor field $T\approx0$ vanish.
Thus $C_i\approx0$ are the first-class constraints.

If the Lagrangian contains no $\dot{N}$, generally the theory has three DoF due to the breaking of
spacetime diffeomorphism \cite{Gao:2014soa}.
When $\dot{N}$ enters the Lagrangian, the lapse function $N$ contributes a scalar mode and generally the theory has four DoF.
 In Ref. \cite{Gao:2018znj}, the authors derived two sufficient and necessary conditions
to eliminate a scalar mode.
The two conditions are
\begin{equation}
\label{DoF3condition1}
\begin{split}
 \mathcal{D}(\vec x,\vec y)=&
  -\int d^3x'\int d^3y'\frac{\delta^2 S_B}{\delta A(\vec x)\delta B_{ij}(\vec x')}
  \mathcal{G}_{ijkl}(\vec x',\vec y')\frac{\delta^2S_B}{\delta B_{kl}(\vec y')\delta A(\vec y)}
  +\frac{\delta^2 S_B}{\delta A(\vec x)\delta A(\vec y)}=0,
  \end{split}
\end{equation}
and
\begin{equation}\label{DoF3condition2}
\begin{split}
  \mathcal{F}  (\vec x,\vec y)=&\int d^3z \{\mathcal{V}_{ij}(\vec z,\vec y)[\tilde{\pi}(\vec x),\tilde{\pi}^{ij}(\vec z)]
   -(\vec x \leftrightarrow \vec y)\}\\
  &+\int d^3x'\int d^3y'
  \mathcal{V}_{ij}(\vec x',\vec x)[\tilde{\pi}^{ij}(\vec x'),\tilde{\pi}^{kl}(\vec y')]
 \mathcal{V}_{kl}(\vec y',\vec y)
  +[\tilde{\pi}(\vec x),\tilde{\pi}(\vec y)]=0,
  \end{split}
\end{equation}
where $\mathcal{G}_{ijkl}(\vec z,\vec x)$ is the inverse of
$\delta^2S_B/\delta B_{ij}(\vec x)\delta B_{kl}(\vec y)$, i.e.,
\begin{equation}
  \int d^3x \mathcal{G}_{ijmn}(\vec z,\vec x)
  \frac{\delta^2S_B}{\delta B_{mn}(\vec x)\delta B_{kl}(\vec y)}=
  \delta^{k}_{i}\delta^l_j\delta^3(\vec z-\vec y),
\end{equation}
and
\begin{equation}
\mathcal{V}_{ij}(\vec x,\vec y)=-2N(\vec x)\int
d^3y'\mathcal{G}_{ijkl}(\vec x,\vec y')[p^{kl}(\vec y'),\tilde{\pi}(\vec y)].
\end{equation}
After imposing the two conditions \eqref{DoF3condition1} and \eqref{DoF3condition2}, another secondary constraint
\begin{equation}\label{C'}
  \mathcal{C}'(\vec x)= [\tilde{\pi}(\vec x),H_c]
  +\int d^3y[\tilde{\pi}^{ij}(\vec y),H_c]\mathcal{V}_{ij}(\vec y,\vec x)\approx0
\end{equation}
arises.
For the sake of counting the number of the physical DoF,
we show the Dirac matrix of the constraints in Table \ref{tab:1}, where the nonzero elements are indicated by * and
\begin{gather}
  \bar{p}(\vec x)=p(\vec x)+\int d^3yp^{ij}(\vec y)\mathcal{U}_{ij}(\vec y,\vec x),\\
  \label{barpi}
  \begin{split}
  \bar{\pi}(\vec x)=&\tilde{\pi}(\vec x)+\int d^3yp^{ij}(\vec y)\mathcal{X}_{ij}(\vec y,\vec x)
  +\int d^3y\tilde{\pi}^{ij}(\vec y)
  \mathcal{V}_{ij}(\vec y,\vec x),
  \end{split}\\
  \begin{split}
  \bar{\mathcal{C}}'(\vec x)=&\mathcal{C}'(\vec x)+\int d^3z [\mathcal{S}_{ij}(\vec x,\vec z)p^{ij}(\vec z)
  +\mathcal{T}_{ij}(\vec x,\vec z)\tilde{\pi}^{ij}(\vec z)],\end{split}\\
\{\varphi^I_1\}=\{\pi_i,\bar{p},p^{ij},\bar{\pi},\tilde{\pi}^{ij},C_i,\bar{C}'\},
\end{gather}
with
\begin{gather}
  \mathcal{U}_{ij}(\vec x,\vec y)=\int d^3x'\mathcal{G}_{ijkl}(\vec x,\vec x')
  2N(\vec x')[\tilde{\pi}^{kl}(\vec x'),p(\vec y)],\\
  \begin{split}
  \mathcal{X}_{ij}(\vec x,\vec y)=&\int d^3x'
  \mathcal{G}_{ijkl}(\vec x,\vec x')2N(\vec x')
  \Bigl([\tilde{\pi}^{kl}(\vec x'),\tilde{\pi}(\vec y)]\\
  &+\int d^3y'[\tilde{\pi}^{kl}(\vec x'),\tilde{\pi}^{mn}(\vec y')]
  \mathcal{V}_{mn}(\vec y',\vec y)\Bigr),
  \end{split}\\
  \mathcal{T}_{ij}(\vec x,\vec w)=\int d^3y' 2N(\vec w)\mathcal{G}_{ijkl}(\vec w,\vec y')[\mathcal{C}'(\vec x),p^{kl}(\vec y')],\\
\begin{split}
  \mathcal{S}_{ij}(\vec x,\vec w)=&-\int d^3y' 2N(\vec y')\mathcal{G}_{ijkl}(\vec w,\vec y')\\
  &\quad \times\Bigl([\mathcal{C}'(\vec x),\tilde{\pi}^{kl}(\vec y')]
  +\int d^3z' \mathcal{T}_{i'j'}(\vec x,\vec z')[\tilde{\pi}^{i'j'}(\vec z'),\tilde{\pi}^{kl}(\vec y')]\Bigr).
\end{split}
\end{gather}

\begin{table*}[htp]
  \centering
  \caption{The Dirac matrix of the constraints. The nonzero elements are indicated by *.}
  \begin{tabular}{c|ccccccc|c}
  \hline\hline
   &$\pi_i(\vec y)$&$\bar{p}(\vec y)$&$p^{kl}(\vec y)$  & $\bar{\pi}(\vec{y})$\  & $\tilde{\pi}^{kl}(\vec{y})$\  & $C_i(\vec y)$ &
   $\bar{\mathcal{C}}'(\vec y)$& $[\varphi^I_1(\vec{x}),H_c]$ \\
  \hline
  $\pi_i(\vec x)$& & & & &  & & & $-C_i(\vec x)$\\
  $\bar{p}(\vec x)$& & & & &  & & & 0 \\
  $p^{ij}(\vec{x})$& & & & &$*$ & & &$0$\\
  $\bar{\pi}(\vec{x})$& & & & & & & * &$*$\\
  $\tilde{\pi}^{kl}(\vec{x})$& & & $*$ & & $*$ & & & $*$\\
  $C_i(\vec{x})$ & & & & & & & &0\\
  $\bar{\mathcal{C}}'(\vec x)$ &  & &  & * & & & * & *\\
  \hline\hline
  \end{tabular}
\label{tab:1}
\end{table*}

From Table \ref{tab:1}, we can see that there are seven first-class
constraints and 14 second-class constraints,
and the physical DoF is
\begin{equation}
  \frac{1}{2}(2n-2n_1-n_2)=(2\times17-2\times7-14)/2=3,
\end{equation}
where $n$, $n_1$, and $n_2$ are the number of canonical variables, first-class constraints and second-class constraints, respectively.

Now let us derive the sufficient and necessary conditions for eliminating another  scalar mode.
In Ref. \cite{Gao:2019twq}, the authors obtained two transverse-traceless(TT) conditions for ensuring two tensorial DoF in the case that the lapse function is nondynamic.
The two TT conditions are \cite{Gao:2019twq}
 \begin{widetext}
 \begin{equation}\label{tt1n}
  \begin{split}
  \mathcal{S}(\vec x,\vec y)=&\frac{\delta^2 S_B}{\delta N(\vec x)\delta N(\vec y)}-
  \int d^3x'\int d^3y'N(\vec x')\frac{\delta}{\delta N(\vec x)}
  \left(\frac{1}{N(\vec x')}\frac{\delta S_B}{\delta B_{i'j'}(\vec x')}\right)\\
  &\qquad\qquad\qquad \qquad \times
   \mathcal{G}_{i'j'k'l'}(\vec x',\vec y')
    N(\vec y')\frac{\delta}{\delta N(\vec y)}
  \left(\frac{1}{N(\vec y')}\frac{\delta S_B}{\delta B_{k'l'}(\vec y')}\right)\approx 0,
\end{split}
\end{equation}
and
\begin{equation}\label{tt2n}
\begin{split}
 \mathcal{J}(\vec x,\vec y)=&\int d^3y'\int d^3y''\mathcal{G}_{ijkl}(\vec y',\vec y'') \left[N(\vec y'')\frac{\delta C'(\vec x)}{\delta h_{kl}(\vec y'')}
  \frac{\delta C'(\vec y)}{\delta B_{ij}(\vec y')}\right.\\
  &\left.+\int d^3y'''\int d^3z' N(\vec z')\mathcal{G}_{i'j'k'l'}(\vec z',\vec y''')
  \frac{\delta C'(\vec x)}{\delta B_{k'l'}(\vec y''')}
  \frac{\delta^{2} S_B}{\delta h_{i'j'}(\vec{z}') \delta B_{kl}(\vec{y}'')}
  \frac{\delta C'(\vec y)}{\delta B_{ij}(\vec y')}\right]\\
  &-(\vec x \leftrightarrow
  \vec y)\approx0,
  \end{split}
\end{equation}
\end{widetext}
where
\begin{equation}\label{c'reduced}
   C'(\vec x)=\frac{\delta S_B}{\delta N}-\frac{1}{N(\vec x)}
   \frac{\delta S_B}{\delta B_{ij}(\vec x)}B_{ij}(\vec x).
\end{equation}
In our case where $N$ becomes dynamical, the two TT conditions \eqref{tt1n} and \eqref{tt2n} need to be generalized.
To eliminate another scalar mode, the Dirac matrix in Table \ref{tab:1}  must be degenerate.
Deleting the first two columns and rows of the Dirac matrix
whose elements are zero, the submatrix is
\begin{equation}
\bm{D}(\vec x,\vec y)=
  \begin{pmatrix}
  0 & \bm{D}_1(\vec x,\vec y)\\
  -\bm{D}^T_1(\vec y,\vec x) & \bm{D}_2(\vec x,\vec y)\\
  \end{pmatrix},
\end{equation}
where
\begin{equation}
  \bm{D}_1(\vec x,\vec y)=
  \left(
  \begin{array}{cc}
  [p^{ij}(\vec x),\tilde{\pi}^{kl}(\vec y)]&0\\
  0&[\bar{\pi}(\vec x),\bar{\mathcal{C}}'(\vec y)]\\
  \end{array}
  \right),
\end{equation}
and
\begin{equation}
  \bm{D}_2(\vec x,\vec y)=
  \left(
  \begin{array}{cc}
 [\tilde{\pi}^{ij}(\vec x),\tilde{\pi}^{kl}(\vec y)] &0\\
   0&[\bar{\mathcal{C}}'(\vec x),\bar{\mathcal{C}}'(\vec y)]\\
  \end{array}
  \right).
\end{equation}
Note that the degeneracy of $\bm{D}$ is equivalent to the degeneracy of $\bm{D}_1$.
To keep the two tensorial degrees,
the degeneracy condition of $\bm{D}_1$ is
\begin{equation}\label{tt1}
  \mathcal{S}_1(\vec x,\vec y)\equiv[\bar{\pi}(\vec x),\bar{\mathcal{C}}'(\vec y)]\approx0.
\end{equation}
Using the explicit expressions of the Poisson brackets presented in  Appendix \ref{app.A}, we get
\begin{equation}\label{tt1expand}
  \begin{split}
  \mathcal{S}_1(\vec x,\vec y)
  &\approx-\frac{\delta \mathcal{C}'(\vec y)}{\delta N(\vec x)}-
  \int d^3y' \mathcal{X}_{ij}(\vec y',\vec x)
  \frac{\delta \mathcal{C}'(\vec y)}{\delta B_{ij}(\vec y')}
  -
  \int d^3y' \mathcal{V}_{ij}(\vec y',\vec x)\frac{\delta \mathcal{C}'(\vec y)}{\delta h_{ij}(\vec y')}.
\end{split}
\end{equation}
Our condition \eqref{tt1} can be regarded as the extension of the first TT condition \eqref{tt1n}.
In fact, when the Lagrangian does not contain $\dot{N}$, $\mathcal{C}'(\vec x)$ reduces to $C'(\vec x)$
and
\begin{equation}
\label{c'red1}
\begin{split}
\mathcal{V}_{ij}&=0,\\
\mathcal{X}_{ij}&=\int d^3x'\mathcal{G}_{ijkl}(\vec x,\vec x')
  2N(\vec x')[\tilde{\pi}^{kl}(\vec x'),\pi(\vec y)],
\end{split}
\end{equation}
with
\begin{equation}
\begin{split}
  [\tilde{\pi}^{kl}(\vec y),\pi(\vec x)]
  =&\frac{1}{2}\delta^3(\vec x-\vec y)\frac{1}{N^2(\vec y)}
  \frac{\delta S_B}{\delta B_{kl}(\vec y)}
  -\frac{1}{2}\frac{1}{N(\vec y)}
  \frac{\delta^2 S_B}{\delta N(\vec x) \delta B_{kl}(\vec y)}.
  \end{split}
\end{equation}
Substituting Eqs. \eqref{c'reduced} and \eqref{c'red1} into Eq. \eqref{tt1expand},
we find our condition \eqref{tt1} reduces to the first TT condition \eqref{tt1n}.

If $[\bar{\mathcal{C}}'(\vec x),\bar{\mathcal{C}}'(\vec y)]\neq0$, then the condition \eqref{tt1} makes $\bar{\pi}\approx0$ a first-class constraint
and no secondary constraint appears.
In this case, we have
eight first-class constraints $\pi_i$, $C_i$, $\bar{p}$ and $\bar{\pi}$,
and 13 second-class constraints $p^{ij}$, $\tilde{\pi}^{ij}$ and $\bar{\mathcal{C}}'$.
The dimension of phase space is
$2\times17-2\times8-13=5$, which is odd and the theory is
not self-consistent.
Thus another condition
\begin{equation}\label{tt2}
  \mathcal{J}_1(\vec x,\vec y)\equiv[\bar{\mathcal{C}}'(\vec x),\bar{\mathcal{C}}'(\vec y)]\approx0
\end{equation}
should be imposed.
Using the explicit expressions of the Poisson brackets in Appendix \ref{app.A}, we get
\begin{equation}
\begin{split}
  &\mathcal{J}_1(\vec x,\vec y)\\
  \approx&\int d^3y'\int d^3y''\mathcal{G}_{ijkl}(\vec y',\vec y'') \left[ 2N(\vec y'')\frac{\delta \mathcal{C}'(\vec x)}{\delta h_{kl}(\vec y'')}
  \frac{\delta \mathcal{C}'(\vec y)}{\delta B_{ij}(\vec y')}\right.\\
  &\left.+\int d^3y'''\int d^3z'2N(\vec z')
  \mathcal{G}_{i'j'k'l'}(\vec z',\vec y''')\frac{\delta \mathcal{C}'(\vec x)}{\delta B_{k'l'}(\vec y''')}
  \frac{\delta^{2} S_B}{\delta h_{i'j'}(\vec{z}') \delta B_{kl}(\vec{y}'')}
  \frac{\delta \mathcal{C}'(\vec y)}{\delta B_{ij}(\vec y')}\right]-(\vec x \leftrightarrow
  \vec y).\\
  \end{split}
\end{equation}
Our condition \eqref{tt2} can be regarded as the extension of the second TT condition \eqref{tt2n}.
In fact, when the Lagrangian does not contain $\dot{N}$, $\mathcal{C}'(\vec x)$ reduces to $C'(\vec x)$ and our condition \eqref{tt2} becomes the second TT condition \eqref{tt2n}.

In the case that $\mathcal{C}'(\vec x)$ only depends on $B_{ij}$, $R_{ij}$, $N$ and $h_{ij}$, i.e. $\mathcal{C}'=\mathcal{C}'(B_{ij},R_{ij},N,h_{ij})$, the second condition \eqref{tt2} can be simplified.
After introducing test functions $\alpha(\vec x)$ and $\beta(\vec y)$ with $\alpha,\beta\xrightarrow{\vec x\rightarrow \infty}0$,
the second   condition \eqref{tt2} becomes
\begin{equation}\label{tt2'}
\begin{split}
 &\int d^3x\frac{2}{\sqrt{h}}\left\{
  (D_n\alpha D_m\beta-D_n\beta D_m\alpha)\Theta_{kl}T^{mnkl}_{ij}\Omega^{ij}\right.\\
  &\left.\quad+
  [T^{mnkl}_{ij}\Omega^{ij}D_n\Theta_{kl}-\Theta_{kl}D_n(T^{nmkl}_{ij}\Omega^{ij})]
  (\alpha D_m\beta-\beta D_m\alpha)
  \right\}  \approx 0,
  \end{split}
\end{equation}
where
\begin{gather}
T^{mnkl}_{ij}=h^{m(l}h^{k)}_{(i}h^n_{j)}-\frac{1}{2}h^{mn}h^{(k}_ih^{l)}_j
-\frac{1}{2}h^{kl}h^m_ih^n_j,\\
\Theta_{kl}=\bar{\mathcal{G}}_{ijkl}\frac{\partial \mathcal{C}'}{\partial B_{ij}},\
\Omega^{ij}=\frac{\partial \mathcal{C}'}{\partial R_{ij}},
\end{gather}
with
\begin{gather}
\bar{\mathcal{G}}_{ijkl}\delta^3(\vec x-\vec y)=N\sqrt{h}
\mathcal{G}_{ijkl}(\vec x,\vec y).
\end{gather}
In the case that $\mathcal{C}'(\vec x)$ depends on the Ricci tensor $R_{ij}$ only through the Ricci scalar $R$, we have
\begin{equation}
  T^{mnkl}_{ij}\Omega^{ij}\propto
  h^{n(k}h^{l)m}-h^{mn}h^{kl},
\end{equation}
so $T^{mnkl}_{ij}\Omega^{ij}$ possesses $m\leftrightarrow n$ exchange symmetry.
The second condition \eqref{tt2'} can be further simplified as
\begin{equation}\label{sttcondition}
\begin{split}
  \int d^3x&\frac{2}{\sqrt{h}}[
  (\alpha D_m\beta-\beta D_m\alpha)
  (T^{mnkl}_{ij}\Omega^{ij}D_n\Theta_{kl}
  -\Theta_{kl}D_nT^{mnkl}_{ij}\Omega^{ij})]\approx0.
  \end{split}
\end{equation}
After the two   conditions \eqref{tt1} and \eqref{tt2} are imposed, a tertiary constraint
\begin{equation}
  \Phi(\vec x)=[\bar{\mathcal{C}}'(\vec x),H_c]
\end{equation}
arises.
In general, we have
\begin{equation}
[\Phi,\bar{p}]\neq0,\
[\Phi,\bar{\pi}]\neq0,\
[\Phi,\bar{\mathcal{C}}']\neq0.
\end{equation}
But we can introduce the combinations
\begin{equation}
    \tilde{\bar{p}}(\vec y)=\int d^3z [\mathcal{R}_1(\vec y,\vec z)\bar{p}(\vec z)+\mathcal{R}_2(\vec y,\vec z)\bar{\pi}(\vec z)],
\end{equation}
and
\begin{equation}
    \tilde{\bar{\mathcal{C}}}'(\vec y)=\int d^3z [\mathcal{R}_3(\vec y,\vec z)\bar{\mathcal{C}}'(\vec z)+\mathcal{R}_4(\vec y,\vec z)\bar{\pi}(\vec z)],
\end{equation}
such that
\begin{equation}
    [\tilde{\bar{p}},\Phi]=0,\
    [\tilde{\bar{\mathcal{C}}}',\Phi]=0.
\end{equation}
Now we have 8 first-class constraints $\{\pi_i,\tilde{\bar{p}},C_i,
\tilde{\bar{\mathcal{C}}}'\}$ and
14 second-class constraints $\{p^{ij},\bar{\pi},\tilde{\pi}^{ij},
\Phi\}$, thus the theory has DoF=$(2\times17-2\times8-14)/2=2$.
We show the Dirac matrix of these constraints in Table \ref{tab:2},
where $\{\varphi^I_2\}=\{\pi_i,\tilde{\bar{p}},p^{ij},\bar{\pi},\tilde{\pi}^{ij},C_i,\tilde{\bar{C}}',\Phi\}$.
\begin{table*}[htp]
  \centering
  \caption{The Dirac matrix of the constraints. The nonzero elements are indicated by *.}
  \begin{tabular}{c|cccccccc|c}
  \hline\hline
   &$\pi_i(\vec y)$&$\tilde{\bar{p}}(\vec y)$&$p^{kl}(\vec y)$  & $\bar{\pi}(\vec{y})$\  & $\tilde{\pi}^{kl}(\vec{y})$\  & $C_i(\vec y)$ &
   $\tilde{\bar{\mathcal{C}}}'(\vec y)$& $\Phi(\vec y)$&$[\varphi_2^I(\vec{x}),H_c]$ \\
  \hline
  $\pi_i(\vec x)$& & & & &  & & & & $-C_i(\vec x)$\\
  $\tilde{\bar{p}}(\vec x)$& & & & &  & & & & 0 \\
  $p^{ij}(\vec{x})$& & & & &$*$ & & & * &$0$\\
  $\bar{\pi}(\vec{x})$& & & & & & &  & * &$*$\\
  $\tilde{\pi}^{kl}(\vec{x})$& & & $*$ & & $*$ & & &* & $*$\\
  $C_i(\vec{x})$ & & & & & & & & &0\\
  $\tilde{\bar{\mathcal{C}}}'(\vec x)$ &  & &  &  & & & & & *\\
  $\Phi(\vec x)$ &  &  &  *& * &* & &  & *& *\\
  \hline\hline
  \end{tabular}
\label{tab:2}
\end{table*}

\section{Quadratic action and field transformation}\label{sec3}

In this section, we construct a class of quadratic actions with two DoF in the case that the lapse function is
dynamic.
 Let us start from the action
\begin{equation}\label{quadraaction1}
\begin{split}
  S=\int d^4xN\sqrt{h}[aK_{ij}K^{ij}+\beta K^2+c_1KF
  +c_2F^2+\gamma\ ^{(3)}R+f],
  \end{split}
\end{equation}
where $^{(3)}R$ is the spatial curvature.
As a first step, we assume that the coupling functions $a$, $\beta$, $c_1$, $c_2$, $\gamma$
and $f$ only depend on $N$.
Generally, this theory has four DoF.
Applying the conditions \eqref{DoF3condition1} and \eqref{DoF3condition2}
to eliminate a scalar mode, we get
\begin{equation}
\label{ivcond1}
  c_2=\frac{3}{4}\frac{c_1^2}{a+3\beta}.
\end{equation}
Note that there is no constraint on $\gamma$ and $f$.
Substituting this condition \eqref{ivcond1} into the action \eqref{quadraaction1}, we get
\begin{equation}\label{qudra}
\begin{split}
   S=\int d^4x N\sqrt{h}[a\hat{K}_{i j} \hat{K}^{i j}+b(K+c F)^{2}+\gamma\ ^{(3)}R+f],
    \end{split}
\end{equation}
where the traceless part of the extrinsic curvature is
\begin{equation}
  \hat{K}_{ij}=K_{ij}-\frac{1}{3}Kh_{ij},
\end{equation}
and
\begin{equation}
  b=\frac{1}{3}(a+2\beta),\
  c=\frac{c_1}{2b}.
\end{equation}
Substituting the equivalent action
\begin{equation}\label{quadraactionlagrangian}
\begin{split}
     S=S_B+
  \int dtd^3x\left[\frac{\delta S_B}{\delta B_{ij}}(K_{ij}-B_{ij})+\frac{\delta  S_B}{\delta A}(F-A)\right],
    \end{split}
\end{equation}
into Eq. \eqref{C'}, we get
\begin{equation}\label{c1}
\begin{split}
  \mathcal{C}'(\vec{x}) =\sqrt{h}[\bar a\hat{B}_{ij}\hat{B}^{ij}
  +\bar b(B+cA)^2+\bar \gamma\ ^{(3)}R+\bar{f}_1+\bar f_2],
 \end{split}
 \end{equation}
 where
\begin{equation}
  \hat{B}_{ij}=B_{ij}-\frac{1}{3}Bh_{ij}
\end{equation}
is the traceless part of $B_{ij}$,
 \begin{equation}
 \begin{split}
   &\bar a=Na'-a-Nac,\
   \bar b=Nb'-b-Nbc,\\
   &\bar \gamma=(N\gamma)'-c\gamma N/3,\
   \bar f_1=(Nf)'-Ncf,\
   \bar f_2=\frac{4c}{3}D_kD^k(\gamma N),
   \end{split}
 \end{equation}
 and the prime denotes the derivative with respect to  $N$.

 Plugging Eq. \eqref{c1} into the first   condition \eqref{tt1}, using the constraint $\mathcal{C}'(\vec x)\approx0$ and introducing the test function $\alpha(\vec y)$, the first condition becomes
\begin{equation}
\begin{split}
  Q_1D^kD_k\alpha+Q_2^kD_k\alpha+Q_3\alpha\approx0,
\end{split}
\end{equation}
where
\begin{equation}
  \begin{split}
  &Q_1=\frac{4}{3}((\gamma N)'c+c\bar \gamma),\\
  &Q_2^k=\frac{8}{3}((\gamma N)'c'+c\bar \gamma')D^kN,\\
  &Q_3=\frac{4}{3}[P_1D_kD^kN+P_2D^kND_kN+P_3]
  \end{split}
\end{equation}
with
\begin{equation}
\begin{split}
  P_1=&(c'-4c^2/3)(\gamma N)'+c\bar \gamma'+(\gamma N)'c',\\
  P_2=&(c'-4c^2/3)(\gamma N)''+c\bar \gamma''+(\gamma N)'c'',\\
  P_3=&(\bar a'-2\bar aa'/a)\hat{B}_{ij}\hat{B}^{ij}+
  \left(\bar b'-2b'\bar b/b\right)\left(B+c A\right)^2
  +(\bar \gamma'-4\bar \gamma c/3)\ ^{(3)}R+\bar f_1'-2c\bar f_1.\\
  \end{split}
\end{equation}
Thus the solution to the first  condition \eqref{tt1} is
\begin{equation}\label{s1concrete}
\begin{split}
  &Q_1=0,\ Q_2^k=0,\ P_1=0,\ P_2=0,P_3=0.
  \end{split}
\end{equation}

In the case that $N$ is dynamic, i.e. $c\neq0$, solving Eq. \eqref{s1concrete}, we get
\begin{equation}\label{cneq0para}
  \begin{split}
   a&=\frac{\alpha_1(t)N}{\alpha_2(t)+N},\
   b=\frac{\alpha_3(t)N}{\alpha_4(t)+N},\\
   c&=-\frac{3}{4N+\alpha_5(t)},\ \gamma=0,
  \end{split}
\end{equation}
where $\alpha_i(t)$ are general functions of time.
The solution $\gamma=0$ indicates that the spatial curvature
 term $^{(3)}R$ does not enter the Lagrangian and
this theory possesses no wave solution and cannot recover GR, so we are not interested in this case.

In the case that $N$ is nondynamic, i.e. $c=0$, the quadratic action was constructed  in  Ref. \cite{Gao:2019twq} as
\begin{equation}\label{quadraactionnoNdot}
\begin{split}
   S=&\int d^4x N\sqrt{h}\left[\frac{N}{\beta_2+N}\hat{K}_{ij}\hat{K}^{ij}-\frac{2N}{3(\beta_4+N)}K^2
  +\left(\beta_5+\frac{\beta_6}{N}\right)\ ^{(3)}R+\beta_7+\frac{\beta_8}{N}\right],
   \end{split}
\end{equation}
where $\beta_i(t)$ are arbitrary functions of time.

Note that throughout the above analysis, we assume that the coupling functions only depend on $N$.
One may ask what if the coupling functions also depend on the spatial derivative of $N$.
However, in this case, the two conditions \eqref{tt1} and \eqref{tt2} become very difficult to solve.
Fortunately, it was proved that an invertible field transformation does not change the number of physical DoF \cite{Arroja:2015wpa,Domenech:2015tca,Takahashi:2017zgr}.
Thus, we take advantage of this result and perform a conformal transformation on the quadratic action \eqref{quadraactionnoNdot} to obtain a new action that contains functions
of $N$, $\dot N$ and $D_i N$. For this purpose, we choose the following conformal transformation
\begin{equation}
\label{conftrans1}
  h_{ij}\rightarrow \mathrm{e}^{2w(N)}h_{ij},\ N\rightarrow N,\ N^i\rightarrow N^i,
\end{equation}
where $w(N)$ is an arbitrary function of $N$.
 Under the conformal transformation \eqref{conftrans1}, the spatial curvature becomes
 \begin{equation}
^{(3)}R \rightarrow \mathrm{e}^{-2w}[^{(3)}R-2D_kwD^kw-4D^kD_kw].
 \end{equation}
The traceless part and the trace of the extrinsic curvature become
\begin{equation}
  \hat{K}_{ij}\rightarrow \mathrm{e}^{2w}\hat{K}_{ij},\
  K\rightarrow K+3w'\mathsterling_{\vec n}N.
\end{equation}
In the case that $w'\neq0$, the lapse function becomes dynamic after the conformal transformation and the corresponding quadratic action is
\begin{equation}\label{quadraticactionmain}
\begin{split}
  S=&\int dtd^3xN\sqrt{h}\left[
 \tilde{a}\hat{K}_{ij}\hat{K}^{ij}
  +\tilde{b}(K+cF)^2+\tilde{\gamma}\ ^{(3)}R+\tilde{f}_1D_kN
  D^kN+\tilde{f}_2D_kD^kN+\tilde{g}\right],\\
  \end{split}
\end{equation}
where
\begin{equation}\label{cf}
\begin{split}
  &\tilde{a}=\frac{\mathrm{e}^{3w}N}{\beta_2+N},\
  \tilde{b}=-\frac{2\mathrm{e}^{3w}N}{3(\beta_1+N)},\
  \tilde{c}=3w',\\
  &\tilde{\gamma}=\mathrm{e}^{w}\left(\beta_5+\frac{\beta_6}{N}\right),\
  \tilde{g}=\mathrm{e}^{3w}\left(\beta_7+\frac{\beta_8}{N}\right),\\
  &\tilde{f}_1=-\tilde{\gamma}(2w'^2+4w''),\
  \tilde{f}_2=-4w'\tilde{\gamma}.
  \end{split}
\end{equation}

To confirm that this theory really propagates two DoF, we check whether the
two   conditions \eqref{tt1} and \eqref{tt2} are satisfied.
After tedious calculations, the first   condition \eqref{tt1} becomes
\begin{equation}\label{s'}
 \bar{Q}_1\alpha+\bar{Q}_2^kD_k\alpha+\bar{Q}_3D^kD_k\alpha\approx0,
\end{equation}
where $\alpha(\vec x)$ is a test function and
\begin{gather}
\begin{split}
   &\bar Q_1= \bar{\tilde{f}}_2+\frac{4\tilde{c}}{3}\bar{\tilde{\gamma}},\\
   &\bar Q_2^k=(-\bar{\tilde{f}}_1+\bar{\tilde{f}}_2'+\frac{4\tilde{c}}{3}\bar{\tilde{\gamma}}'
   +\frac{\tilde{c}}{6}\bar{\tilde{f}}_2)D^kN,\\
 &\bar Q_3=\bar P_1+\bar P_2D^kD_kN+\bar P_3D_kND^kN,
 \end{split}
   \end{gather}
 with
 \begin{equation}
 \begin{split}
 &\bar{\tilde{\gamma}}=(\tilde{\gamma}N)'-\frac{1}{3}\tilde{c}\tilde{\gamma}N,\\
 &\bar{\tilde{f}}_1=-(\tilde{f}_1N)'+(\tilde{f}_2N)''-\frac{1}{3}N\tilde{c}
   \tilde{f}_1
   +\frac{4\tilde{c}}{3}(\tilde{\gamma}N)''+\frac{\tilde{c}}{3}(N\tilde{f}_2)',\\
 &\bar{\tilde{f}}_2=2(\tilde{f}_2N)'-2\tilde{f}_1N-\frac{1}{3}N\tilde{c}\tilde{f}_2
   +\frac{4\tilde{c}}{3}(\tilde{\gamma}N)'+\frac{\tilde{c}}{3}N\tilde{f}_2,\\
  &\bar P_1=(\bar{\tilde{a}}'-2\bar{\tilde{a}}\tilde{a}'/\tilde{a}+c\bar{\tilde{a}})
  \hat{B}_{ij}\hat{B}^{ij}+
  (\bar{\tilde{b}}'-2\tilde{b'}\bar{\tilde{b}}/\tilde{b}+\tilde{c}\bar{\tilde{b}})\\
&\ \ \ \ \ \times(B+\tilde{c} A)^2+(\bar{\tilde{\gamma}}'-\bar{\tilde{\gamma}}\tilde{c}/3)^{(3)}R+\bar{\tilde{g}}'-\tilde{c}
\bar{\tilde{g}},\\
  &\bar P_2=\bar{\tilde{f}}'_2-\tilde{c}\bar{\tilde{f}}_2/3-2\bar{\tilde{f}}_1
  +\bar{\tilde{f}}_2'
  +\frac{4}{3}\tilde{c}\bar{\tilde{\gamma}}'+\frac{\tilde{c}}{3}\bar{\tilde{f}}_2,\\
  &\bar P_3=\bar{\tilde{f}}'_1-\tilde{c}\bar{\tilde{f}}_1/3-2\bar{\tilde{f}}_1'
  +\bar{\tilde{f}}_2''
  +\frac{4}{3}\tilde{c}\bar{\tilde{\gamma}}''+\frac{\tilde{c}}{3}\bar{\tilde{f}}_2'.
  \end{split}
\end{equation}
and
 \begin{equation}
 \begin{split}
&\bar{\tilde{a}}=N\tilde{a}'-\tilde{a}-N\tilde{a}\tilde{c},\ \bar{\tilde{b}}=N\tilde{b}'-\tilde{b}-N\tilde{b}\tilde{c},\
 \bar{\tilde{g}}=(N\tilde{g})'-N\tilde{c}\tilde{g}.\\
   \end{split}
\end{equation}
Using the explicit expression of the coupling functions $\tilde{a}$, $\tilde{b}$, $\tilde{c}$,
 $\tilde{\gamma}$, $\tilde{f}_1$, $\tilde{f}_2$ and $\tilde{g}$ from Eq. \eqref{cf}, we find that
\begin{equation}
  \bar Q_1=0,\ \bar Q^k_2=0,\ \bar Q_3=0,
\end{equation}
and thus the first   condition $\mathcal{S}_1(\vec x,\vec y)\approx0$ is satisfied. The second   condition \eqref{tt2} is
\begin{equation}
\label{tt2sol5}
\begin{split}
&\frac{1}{\sqrt{h}}(T^{mnkl}_{ij}\Omega^{ij}D_n\Theta_{kl}-
\Theta_{kl}D_nT^{mnkl}_{ij}\Omega^{ij})\\
  =&\sqrt{h}\left(\bar{\tilde{\gamma}}D_n\left[\frac{\bar{\tilde{a}}}{\tilde{a}}B^{mn}
  +\left(-\frac{\bar{\tilde{a}}}{3\tilde{a}}-\frac{2\bar{\tilde{b}}}{3\tilde{b}}\right)
  Bh^{mn}-\frac{2\bar{\tilde{b}}}{3\tilde{b}}\tilde{c}Ah^{mn}\right]\right.\\
 & \left. -\left[\frac{\bar{\tilde{a}}}{\tilde{a}}B^{mn}
  +\left(-\frac{\bar{\tilde{a}}}{3\tilde{a}}-\frac{2\bar{\tilde{b}}}{3\tilde{b}}\right)
  Bh^{mn}-\frac{2\bar{\tilde{b}}}{3\tilde{b}}\tilde{c}Ah^{mn}\right]D_n\bar{\tilde{\gamma}}\right).
 \end{split}
\end{equation}
Using the explicit expressions of the coupling functions $\tilde{a}$, $\tilde{b}$ and $\tilde{c}$ from Eq. \eqref{cf}, we find the following relations
\begin{equation}
\label{tt2sol6}
\begin{split}
  &\frac{\bar{\tilde{a}}}{\tilde{a}}=-\mathrm{e}^{-3w}\tilde{a},\
  -\frac{\bar{\tilde{a}}}{3\tilde{a}}-\frac{2\bar{\tilde{b}}}{3\tilde{b}}=-\mathrm{e}^{-3w}
  (\tilde{b}-\tilde{a}/3),\
-\frac{2\bar{\tilde{b}}}{3\tilde{b}}\tilde{c}=-\mathrm{e}^{-3w}\tilde{b}\tilde{c}.
 \end{split}
\end{equation}
Substituting Eq. \eqref{tt2sol6} into Eq. \eqref{tt2sol5}, we get
\begin{equation}
  \begin{split}
&\frac{1}{\sqrt{h}}(T^{mnkl}_{ij}\Omega^{ij}D_n\Theta_{kl}-
\Theta_{kl}D_nT^{mnkl}_{ij}\Omega^{ij})\\
  =&-\sqrt{h}\left(\bar{\tilde{\gamma}}D_n \left[\mathrm{e}^{-3w}\left(\tilde{a}B^{mn}
  +(\tilde{b}-\tilde{a}/3)
  Bh^{mn}+\tilde{b}\tilde{c}Ah^{mn}\right)\right]\right.\\
&\left. -\mathrm{e}^{-3w}\left(\tilde{a}B^{mn}
  +(\tilde{b}-\tilde{a}/3)Bh^{mn}+\tilde{b}\tilde{c}Ah^{mn}\right)D_n\bar{\tilde{\gamma}}\right).\\
 \end{split}
\end{equation}
Applying the constraint
\begin{equation}
\begin{split}
  C^i(\vec x)\approx  -2\sqrt{h}D_j\left(
  \tilde{a}B^{ij}+(\tilde{b}-\tilde{a}/3)Bh^{ij}+\tilde{b}\tilde{c}Ah^{ij}\right)
  \approx0,\\
  \end{split}
\end{equation}
and choosing the Minkowski spacetime as the boundary condition at infinity,
 we have
\begin{equation}
  \tilde{a}B^{ij}+(\tilde{b}-\tilde{a}/3)Bh^{ij}+\tilde{b}\tilde{c}Ah^{ij}\approx0.
\end{equation}
Therefore, the second   condition \eqref{tt2} is also satisfied and we confirm that the theory \eqref{quadraticactionmain} has only two propagating DoF.

Motivated by the field transformation, one may wonder
whether there exists the correspondence
between GR and  SCG with two DoF.
Under the disformal transformation
\begin{equation}
  h_{ij}\rightarrow \mathrm{e}^{2w(N)}h_{ij},\ N\rightarrow \mathrm{e}^{\lambda(N)}N,\ N^i\rightarrow N^i,
\end{equation}
the action for GR
\begin{equation}
  S_{\mathrm{GR}}=\int d^4x N\sqrt{h}\left[\hat{K}_{ij}\hat{K}^{ij}-\frac{2}{3}K^{2}+\ ^{(3)}R\right]\\
\end{equation}
becomes
\begin{equation}
\begin{split}
S=&\int d^4x N\sqrt{h}\left\{e^{3w-\lambda}\hat{K}_{ij}\hat{K}^{ij}
  -\frac{2}{3}e^{3w-\lambda}(K+3w'F)^{2}\right.\\
  &\left.+e^{w+\lambda}[^{(3)}R-4w'D_kD^kN
  -(2w'^2+4w'')D_kND^kN]\right\},
  \end{split}
 \end{equation}
where $w(N)$ and $\lambda(N)$ are arbitrary functions of $N$.
Choosing
  \begin{equation}
    \lambda(N)=\mathrm{ln}\left(\frac{\gamma_1+\gamma_2N}{N}\right),
  \end{equation}
 we get
 \begin{equation}
 \label{action10}
 \begin{split}
S=&\int d^4x N\sqrt{h}\left\{\frac{\mathrm{e}^{3w}N}{\gamma_1+\gamma_2N}\hat{K}_{ij}\hat{K}^{ij}
  -\frac{2\mathrm{e}^{3w}N}{3(\gamma_1+\gamma_2N)}(K+3w'F)^{2}\right.
\\
  &\left.+\mathrm{e}^{w}\left(\frac{\gamma_1}{N}
  +\gamma_2\right)[^{(3)}R-4w'D_kD^kN
  -(2w'^2+4w'')D_kND^kN]\right\}.
  \end{split}
 \end{equation}
It is obvious that the action \eqref{action10} is a subclass of the quadratic action  \eqref{quadraticactionmain},
so GR can be transformed to SCG with dynamic $N$ by the disformal transformation.
In the case $w=0$, we get
\begin{equation}
\begin{split}
S=&\int d^4x N\sqrt{h}\left[\frac{N}{\gamma_1+\gamma_2N}\hat{K}_{ij}\hat{K}^{ij}
-\frac{2N}{3(\gamma_1+\gamma_2N)}K^{2}+\left(\frac{\gamma_1}{N}+\gamma_2\right)\ ^{(3)}R\right],
 \end{split}
 \end{equation}
 which belongs to  a subclass of SCG quadratic actions with nondynamic $N$ constructed in Ref. \cite{Gao:2019twq}.

 \section{Conclusion}\label{sec4}
 In this paper, within the framework of spatially covariant gravity  with a dynamic lapse function $N$, we investigated the sufficient and necessary conditions for a theory to have two physical degrees of freedom by performing the
 Hamiltonian analysis.
 Generally, the dynamic lapse function contributes a scalar mode and the theory has four DoF.
 In Ref. \cite{Gao:2018znj} two conditions have been obtained to eliminate a scalar mode.
 We further obtained the sufficient and necessary conditions \eqref{tt1} and \eqref{tt2} to eliminate another scalar mode through the detailed Hamiltonian analysis and they are
 $$
  [\bar{\pi}(\vec x),\bar{\mathcal{C}}'(\vec y)]\approx0,\
  [\bar{\mathcal{C}}'(\vec x),\bar{\mathcal{C}}'(\vec y)]\approx0.
$$
The first condition \eqref{tt1} ensures the Dirac matrix is degenerate and turns a second-class constraint to a first-class constraint.
If only the first   condition is imposed, the dimension of phase space at each point of spacetime becomes odd and the theory is not self-consistent.
Thus another condition should be imposed.
The second   condition \eqref{tt2} ensures that a tertiary constraint arises. The dimension of the phase space then becomes even and the theory has two DoF.

In the case that coupling functions only depend on $N$, we found that the spatial curvature term cannot enter the Lagrangian, and the theory possesses no wave solution and cannot recover GR.
To construct SCG with coupling functions dependent on the spatial
derivative of $N$,
we performed a spatially conformal transformation, i.e.
$h_{ij}\rightarrow \mathrm{e}^{2w}h_{ij}$, on a class of quadratic actions with a nondynamic lapse function and two DoF.
Due to the $N$ dependence of spatial conformal factor $w$, lapse function becomes dynamic after the conformal transformation, and a class of quadratic actions with $\dot{N}$ and two DoF is obtained.
We confirmed that this theory propagates two DoF by checking the two  sufficient and necessary  conditions \eqref{tt1} and \eqref{tt2}.
Besides, we also investigated the correspondence between SCG with two DoF and GR.
We found that a subclass of the quadratic actions \eqref{quadraticactionmain} with dynamic $N$  can be related to GR by performing the spatially conformal transformation $h_{ij}\rightarrow \mathrm{e}^{2w}h_{ij}$
and rescaling the lapse function $N\rightarrow e^{2\lambda}N$, with
$\lambda=\ln\left[(\gamma_1+\gamma_2N)/N\right]$.

\begin{acknowledgments}
J. Lin thanks Prof. Xian Gao for useful discussions.
This research was supported in part by the National Natural Science Foundation of
China under Grant No. 11875136 and the Major Program of the National Natural Science
Foundation of China under Grant No. 11690021.
\end{acknowledgments}

  \appendix

\section{The Poisson bracket}\label{app.A}
The Poisson brackets among primary constraints are
\begin{equation}
    [\pi_i(\vec x), \varphi^I(\vec y)]=0,
\end{equation}
\begin{equation}
    [p(\vec x),p(\vec y)]=0,
\end{equation}
\begin{equation}
    [p^{ij}(\vec x),p^{kl}(\vec y)]=0,
\end{equation}
\begin{equation}
  [p(\vec x),\tilde{\pi}(\vec y)]=\frac{1}{N(\vec y)}\frac{\delta^2 S_B}{\delta A(\vec x)\delta A(\vec y)},
\end{equation}
\begin{equation}
  [p(\vec x),\tilde{\pi}^{kl}(\vec y)]=\frac{1}{2N(\vec y)}\frac{\delta^2 S_B}{\delta A(\vec x)\delta B_{kl}(\vec y)},
\end{equation}
\begin{equation}
  [p^{ij}(\vec x),\tilde{\pi}(\vec y)]=\frac{1}{N(\vec y)}\frac{\delta^2 S_B}{\delta B_{ij}(\vec x)\delta A(\vec y)},
\end{equation}
\begin{equation}
  [p^{ij}(\vec x),\tilde{\pi}^{kl}(\vec y)]=\frac{1}{2N(\vec y)}\frac{\delta^2 S_B}{\delta B_{ij}(\vec x)\delta B_{kl}(\vec y)},
\end{equation}
\begin{equation}
  [\tilde{\pi}(\vec x),\tilde{\pi}(\vec y)]=\frac{1}{N(\vec y)}\frac{\delta^2 S_B}{\delta N(\vec x)\delta A(\vec y)}
  -\frac{1}{N(\vec x)}\frac{\delta^2 S_B}{\delta A(\vec x)\delta N(\vec y)},
\end{equation}
\begin{equation}
\begin{split}
  [\tilde{\pi}(\vec x),\tilde{\pi}^{ij}(\vec y)]=&
  -\frac{1}{2}\delta^3(\vec x-\vec y)\frac{1}{N^2(\vec y)}\frac{\delta S_B}{\delta B_{ij}(\vec y)}
  +\frac{1}{2N(\vec y)}\frac{\delta^2 S_B}{\delta N(\vec x)\delta B_{ij}(\vec y)}\\
  &-\frac{1}{N(\vec x)}\frac{\delta^2 S_B}{\delta A(\vec x)\delta h_{ij}(\vec y)},
  \end{split}
\end{equation}
\begin{equation}
  [\tilde{\pi}^{ij}(\vec x),\tilde{\pi}^{kl}(\vec y)]=\frac{1}{2N(\vec y)}\frac{\delta^2 S_B}{\delta h_{ij}(\vec x)\delta B_{kl}(\vec y)}
  -\frac{1}{2N(\vec x)}\frac{\delta^2 S_B}{\delta B_{ij}(\vec x)\delta h_{kl}(\vec y)}.
\end{equation}

The Poisson brackets between primary constraints and the canonical Hamiltonian are
\begin{equation}
[p(\vec x),H_C]\approx 0,
\end{equation}
\begin{equation}
[p^{ij}(\vec x),H_C]\approx 0,
\end{equation}
\begin{equation}
[\pi_i(\vec x),H_C]=-C_i(\vec x),
\end{equation}
\begin{equation}
\begin{split}
[\tilde{\pi}(\vec x),H_C]\approx &
    \frac{\delta S_B}{\delta N(\vec x)}-\frac{1}{N(\vec x)}\frac{\delta S_B}{\delta B_{ij}(\vec x)}B_{ij}(\vec x)\\
    &-\frac{1}{N(\vec x)}\int d^3yN(\vec y)\left(
    \frac{\delta^2 S_B}{\delta A(\vec x)\delta N(\vec y)}+
    \frac{\delta^2S_B}{\delta A(\vec x)\delta h_{ij}(\vec y)}2B_{ij}(\vec y)\right),
    \end{split}
\end{equation}
\begin{equation}
\begin{split}
[\tilde{\pi}^{ij}(\vec x),H_C]=&\frac{\delta S_B}{\delta h_{ij}(\vec x)}
    +\frac{A(\vec x)}{2N(\vec x)}\frac{\delta S_B}{\delta B_{ij}(\vec x)}\\
    &-\frac{1}{2N(\vec x)}\int d^3yN(\vec y)
    \left(\frac{\delta^2S_B}{\delta B_{ij}(\vec x)\delta N(\vec y)}A(\vec y)+\frac{\delta^2S_B}{\delta B_{ij}(\vec x)\delta h_{kl}(\vec y)}2B_{kl}(\vec y)\right).
    \end{split}
\end{equation}


\begin{thebibliography}{71}%
\makeatletter
\providecommand \@ifxundefined [1]{%
 \@ifx{#1\undefined}
}%
\providecommand \@ifnum [1]{%
 \ifnum #1\expandafter \@firstoftwo
 \else \expandafter \@secondoftwo
 \fi
}%
\providecommand \@ifx [1]{%
 \ifx #1\expandafter \@firstoftwo
 \else \expandafter \@secondoftwo
 \fi
}%
\providecommand \natexlab [1]{#1}%
\providecommand \enquote  [1]{``#1''}%
\providecommand \bibnamefont  [1]{#1}%
\providecommand \bibfnamefont [1]{#1}%
\providecommand \citenamefont [1]{#1}%
\providecommand \href@noop [0]{\@secondoftwo}%
\providecommand \href [0]{\begingroup \@sanitize@url \@href}%
\providecommand \@href[1]{\@@startlink{#1}\@@href}%
\providecommand \@@href[1]{\endgroup#1\@@endlink}%
\providecommand \@sanitize@url [0]{\catcode `\\12\catcode `\$12\catcode
  `\&12\catcode `\#12\catcode `\^12\catcode `\_12\catcode `\%12\relax}%
\providecommand \@@startlink[1]{}%
\providecommand \@@endlink[0]{}%
\providecommand \url  [0]{\begingroup\@sanitize@url \@url }%
\providecommand \@url [1]{\endgroup\@href {#1}{\urlprefix }}%
\providecommand \urlprefix  [0]{URL }%
\providecommand \Eprint [0]{\href }%
\providecommand \doibase [0]{https://doi.org/}%
\providecommand \selectlanguage [0]{\@gobble}%
\providecommand \bibinfo  [0]{\@secondoftwo}%
\providecommand \bibfield  [0]{\@secondoftwo}%
\providecommand \translation [1]{[#1]}%
\providecommand \BibitemOpen [0]{}%
\providecommand \bibitemStop [0]{}%
\providecommand \bibitemNoStop [0]{.\EOS\space}%
\providecommand \EOS [0]{\spacefactor3000\relax}%
\providecommand \BibitemShut  [1]{\csname bibitem#1\endcsname}%
\let\auto@bib@innerbib\@empty
\bibitem [{\citenamefont {Guth}(1981)}]{Guth:1980zm}%
  \BibitemOpen
  \bibfield  {author} {\bibinfo {author} {\bibfnamefont {A.~H.}\ \bibnamefont
  {Guth}},\ }\bibinfo {title} {{The Inflationary Universe: A Possible Solution
  to the Horizon and Flatness Problems}},\ \href
  {https://doi.org/10.1103/PhysRevD.23.347} {\bibfield  {journal} {\bibinfo
  {journal} {Phys. Rev. D}\ }\textbf {\bibinfo {volume} {23}},\ \bibinfo
  {pages} {347} (\bibinfo {year} {1981})}\BibitemShut {NoStop}%
\bibitem [{\citenamefont {Starobinsky}(1980)}]{Starobinsky:1980te}%
  \BibitemOpen
  \bibfield  {author} {\bibinfo {author} {\bibfnamefont {A.}~\bibnamefont
  {Starobinsky}},\ }\bibinfo {title} {A new type of isotropic cosmological
  models without singularity},\ \href
  {https://doi.org/https://doi.org/10.1016/0370-2693(80)90670-X} {\bibfield
  {journal} {\bibinfo  {journal} {Phys. Lett. B}\ }\textbf {\bibinfo {volume}
  {91}},\ \bibinfo {pages} {99 } (\bibinfo {year} {1980})}\BibitemShut
  {NoStop}%
\bibitem [{\citenamefont {Linde}(1982)}]{Linde:1981mu}%
  \BibitemOpen
  \bibfield  {author} {\bibinfo {author} {\bibfnamefont {A.~D.}\ \bibnamefont
  {Linde}},\ }\bibinfo {title} {{A New Inflationary Universe Scenario: A
  Possible Solution of the Horizon, Flatness, Homogeneity, Isotropy and
  Primordial Monopole Problems}},\ \href
  {https://doi.org/10.1016/0370-2693(82)91219-9} {\bibfield  {journal}
  {\bibinfo  {journal} {Phys. Lett. B}\ }\textbf {\bibinfo {volume} {108}},\
  \bibinfo {pages} {389} (\bibinfo {year} {1982})}\BibitemShut {NoStop}%
\bibitem [{\citenamefont {Starobinsky}(1982)}]{Starobinsky:1982ee}%
  \BibitemOpen
  \bibfield  {author} {\bibinfo {author} {\bibfnamefont {A.~A.}\ \bibnamefont
  {Starobinsky}},\ }\bibinfo {title} {{Dynamics of Phase Transition in the New
  Inflationary Universe Scenario and Generation of Perturbations}},\ \href
  {https://doi.org/10.1016/0370-2693(82)90541-X} {\bibfield  {journal}
  {\bibinfo  {journal} {Phys. Lett. B}\ }\textbf {\bibinfo {volume} {117}},\
  \bibinfo {pages} {175} (\bibinfo {year} {1982})}\BibitemShut {NoStop}%
\bibitem [{\citenamefont {Riess}\ {\it et~al.}(1998)\citenamefont {Riess} {\it
  et~al.}}]{Riess:1998cb}%
  \BibitemOpen
  \bibfield  {author} {\bibinfo {author} {\bibfnamefont {A.~G.}\ \bibnamefont
  {Riess}} {\it et~al.} (\bibinfo {collaboration} {Supernova Search Team}),\
  }\bibinfo {title} {{Observational evidence from supernovae for an
  accelerating universe and a cosmological constant}},\ \href
  {https://doi.org/10.1086/300499} {\bibfield  {journal} {\bibinfo  {journal}
  {Astron. J.}\ }\textbf {\bibinfo {volume} {116}},\ \bibinfo {pages} {1009}
  (\bibinfo {year} {1998})},\ \Eprint {https://arxiv.org/abs/astro-ph/9805201}
  {arXiv:astro-ph/9805201} \BibitemShut {NoStop}%
\bibitem [{\citenamefont {Perlmutter}\ {\it et~al.}(1999)\citenamefont
  {Perlmutter} {\it et~al.}}]{Perlmutter:1998np}%
  \BibitemOpen
  \bibfield  {author} {\bibinfo {author} {\bibfnamefont {S.}~\bibnamefont
  {Perlmutter}} {\it et~al.} (\bibinfo {collaboration} {Supernova Cosmology
  Project}),\ }\bibinfo {title} {{Measurements of $\Omega$ and $\Lambda$ from
  42 high redshift supernovae}},\ \href {https://doi.org/10.1086/307221}
  {\bibfield  {journal} {\bibinfo  {journal} {Astrophys. J.}\ }\textbf
  {\bibinfo {volume} {517}},\ \bibinfo {pages} {565} (\bibinfo {year}
  {1999})},\ \Eprint {https://arxiv.org/abs/astro-ph/9812133}
  {arXiv:astro-ph/9812133} \BibitemShut {NoStop}%
\bibitem [{\citenamefont {Clifton}\ {\it et~al.}(2012)\citenamefont {Clifton},
  \citenamefont {Ferreira}, \citenamefont {Padilla},\ and\ \citenamefont
  {Skordis}}]{Clifton:2011jh}%
  \BibitemOpen
  \bibfield  {author} {\bibinfo {author} {\bibfnamefont {T.}~\bibnamefont
  {Clifton}}, \bibinfo {author} {\bibfnamefont {P.~G.}\ \bibnamefont
  {Ferreira}}, \bibinfo {author} {\bibfnamefont {A.}~\bibnamefont {Padilla}},\
  and\ \bibinfo {author} {\bibfnamefont {C.}~\bibnamefont {Skordis}},\
  }\bibinfo {title} {{Modified Gravity and Cosmology}},\ \href
  {https://doi.org/10.1016/j.physrep.2012.01.001} {\bibfield  {journal}
  {\bibinfo  {journal} {Phys. Rept.}\ }\textbf {\bibinfo {volume} {513}},\
  \bibinfo {pages} {1} (\bibinfo {year} {2012})},\ \Eprint
  {https://arxiv.org/abs/1106.2476} {arXiv:1106.2476 [astro-ph.CO]}
  \BibitemShut {NoStop}%
\bibitem [{\citenamefont {Horndeski}(1974)}]{Horndeski:1974wa}%
  \BibitemOpen
  \bibfield  {author} {\bibinfo {author} {\bibfnamefont {G.~W.}\ \bibnamefont
  {Horndeski}},\ }\bibinfo {title} {{Second-order scalar-tensor field equations
  in a four-dimensional space}},\ \href {https://doi.org/10.1007/BF01807638}
  {\bibfield  {journal} {\bibinfo  {journal} {Int. J. Theor. Phys.}\ }\textbf
  {\bibinfo {volume} {10}},\ \bibinfo {pages} {363} (\bibinfo {year}
  {1974})}\BibitemShut {NoStop}%
\bibitem [{\citenamefont {Armendariz-Picon}\ {\it et~al.}(1999)\citenamefont
  {Armendariz-Picon}, \citenamefont {Damour},\ and\ \citenamefont
  {Mukhanov}}]{ArmendarizPicon:1999rj}%
  \BibitemOpen
  \bibfield  {author} {\bibinfo {author} {\bibfnamefont {C.}~\bibnamefont
  {Armendariz-Picon}}, \bibinfo {author} {\bibfnamefont {T.}~\bibnamefont
  {Damour}},\ and\ \bibinfo {author} {\bibfnamefont {V.~F.}\ \bibnamefont
  {Mukhanov}},\ }\bibinfo {title} {{k-inflation}},\ \href
  {https://doi.org/10.1016/S0370-2693(99)00603-6} {\bibfield  {journal}
  {\bibinfo  {journal} {Phys. Lett. B}\ }\textbf {\bibinfo {volume} {458}},\
  \bibinfo {pages} {209} (\bibinfo {year} {1999})},\ \Eprint
  {https://arxiv.org/abs/hep-th/9904075} {arXiv:hep-th/9904075} \BibitemShut
  {NoStop}%
\bibitem [{\citenamefont {Garriga}\ and\ \citenamefont
  {Mukhanov}(1999)}]{Garriga:1999vw}%
  \BibitemOpen
  \bibfield  {author} {\bibinfo {author} {\bibfnamefont {J.}~\bibnamefont
  {Garriga}}\ and\ \bibinfo {author} {\bibfnamefont {V.~F.}\ \bibnamefont
  {Mukhanov}},\ }\bibinfo {title} {{Perturbations in k-inflation}},\ \href
  {https://doi.org/10.1016/S0370-2693(99)00602-4} {\bibfield  {journal}
  {\bibinfo  {journal} {Phys. Lett. B}\ }\textbf {\bibinfo {volume} {458}},\
  \bibinfo {pages} {219} (\bibinfo {year} {1999})},\ \Eprint
  {https://arxiv.org/abs/hep-th/9904176} {arXiv:hep-th/9904176} \BibitemShut
  {NoStop}%
\bibitem [{\citenamefont {Kobayashi}\ {\it et~al.}(2010)\citenamefont
  {Kobayashi}, \citenamefont {Yamaguchi},\ and\ \citenamefont
  {Yokoyama}}]{Kobayashi:2010cm}%
  \BibitemOpen
  \bibfield  {author} {\bibinfo {author} {\bibfnamefont {T.}~\bibnamefont
  {Kobayashi}}, \bibinfo {author} {\bibfnamefont {M.}~\bibnamefont
  {Yamaguchi}},\ and\ \bibinfo {author} {\bibfnamefont {J.}~\bibnamefont
  {Yokoyama}},\ }\bibinfo {title} {{G-inflation: Inflation driven by the
  Galileon field}},\ \href {https://doi.org/10.1103/PhysRevLett.105.231302}
  {\bibfield  {journal} {\bibinfo  {journal} {Phys. Rev. Lett.}\ }\textbf
  {\bibinfo {volume} {105}},\ \bibinfo {pages} {231302} (\bibinfo {year}
  {2010})},\ \Eprint {https://arxiv.org/abs/1008.0603} {arXiv:1008.0603
  [hep-th]} \BibitemShut {NoStop}%
\bibitem [{\citenamefont {Crisostomi}\ and\ \citenamefont
  {Koyama}(2018)}]{Crisostomi:2017pjs}%
  \BibitemOpen
  \bibfield  {author} {\bibinfo {author} {\bibfnamefont {M.}~\bibnamefont
  {Crisostomi}}\ and\ \bibinfo {author} {\bibfnamefont {K.}~\bibnamefont
  {Koyama}},\ }\bibinfo {title} {{Self-accelerating universe in scalar-tensor
  theories after GW170817}},\ \href
  {https://doi.org/10.1103/PhysRevD.97.084004} {\bibfield  {journal} {\bibinfo
  {journal} {Phys. Rev. D}\ }\textbf {\bibinfo {volume} {97}},\ \bibinfo
  {pages} {084004} (\bibinfo {year} {2018})},\ \Eprint
  {https://arxiv.org/abs/1712.06556} {arXiv:1712.06556 [astro-ph.CO]}
  \BibitemShut {NoStop}%
\bibitem [{\citenamefont {Crisostomi}\ {\it et~al.}(2019)\citenamefont
  {Crisostomi}, \citenamefont {Koyama}, \citenamefont {Langlois}, \citenamefont
  {Noui},\ and\ \citenamefont {Steer}}]{Crisostomi:2018bsp}%
  \BibitemOpen
  \bibfield  {author} {\bibinfo {author} {\bibfnamefont {M.}~\bibnamefont
  {Crisostomi}}, \bibinfo {author} {\bibfnamefont {K.}~\bibnamefont {Koyama}},
  \bibinfo {author} {\bibfnamefont {D.}~\bibnamefont {Langlois}}, \bibinfo
  {author} {\bibfnamefont {K.}~\bibnamefont {Noui}},\ and\ \bibinfo {author}
  {\bibfnamefont {D.~A.}\ \bibnamefont {Steer}},\ }\bibinfo {title}
  {{Cosmological evolution in DHOST theories}},\ \href
  {https://doi.org/10.1088/1475-7516/2019/01/030} {J. Cosmol. Astropart. Phys.\
  \bibinfo {volume} {01}\bibfield  {year} {\bibinfo  {year} { (\textbf
  {2019})}\ }\bibfield  {pages} {\bibinfo  {pages} {030}},\ }\Eprint
  {https://arxiv.org/abs/1810.12070} {arXiv:1810.12070 [hep-th]} \BibitemShut
  {NoStop}%
\bibitem [{\citenamefont {Frusciante}\ {\it et~al.}(2019)\citenamefont
  {Frusciante}, \citenamefont {Kase}, \citenamefont {Koyama}, \citenamefont
  {Tsujikawa},\ and\ \citenamefont {Vernieri}}]{Frusciante:2018tvu}%
  \BibitemOpen
  \bibfield  {author} {\bibinfo {author} {\bibfnamefont {N.}~\bibnamefont
  {Frusciante}}, \bibinfo {author} {\bibfnamefont {R.}~\bibnamefont {Kase}},
  \bibinfo {author} {\bibfnamefont {K.}~\bibnamefont {Koyama}}, \bibinfo
  {author} {\bibfnamefont {S.}~\bibnamefont {Tsujikawa}},\ and\ \bibinfo
  {author} {\bibfnamefont {D.}~\bibnamefont {Vernieri}},\ }\bibinfo {title}
  {{Tracker and scaling solutions in DHOST theories}},\ \href
  {https://doi.org/10.1016/j.physletb.2019.01.009} {\bibfield  {journal}
  {\bibinfo  {journal} {Phys. Lett. B}\ }\textbf {\bibinfo {volume} {790}},\
  \bibinfo {pages} {167} (\bibinfo {year} {2019})},\ \Eprint
  {https://arxiv.org/abs/1812.05204} {arXiv:1812.05204 [gr-qc]} \BibitemShut
  {NoStop}%
\bibitem [{\citenamefont {Lin}\ {\it et~al.}(2020)\citenamefont {Lin},
  \citenamefont {Gao}, \citenamefont {Gong}, \citenamefont {Lu}, \citenamefont
  {Zhang},\ and\ \citenamefont {Zhang}}]{Lin:2020goi}%
  \BibitemOpen
  \bibfield  {author} {\bibinfo {author} {\bibfnamefont {J.}~\bibnamefont
  {Lin}}, \bibinfo {author} {\bibfnamefont {Q.}~\bibnamefont {Gao}}, \bibinfo
  {author} {\bibfnamefont {Y.}~\bibnamefont {Gong}}, \bibinfo {author}
  {\bibfnamefont {Y.}~\bibnamefont {Lu}}, \bibinfo {author} {\bibfnamefont
  {C.}~\bibnamefont {Zhang}},\ and\ \bibinfo {author} {\bibfnamefont
  {F.}~\bibnamefont {Zhang}},\ }\bibinfo {title} {{Primordial black holes and
  secondary gravitational waves from $k$ and $G$ inflation}},\ \href
  {https://doi.org/10.1103/PhysRevD.101.103515} {\bibfield  {journal} {\bibinfo
   {journal} {Phys. Rev. D}\ }\textbf {\bibinfo {volume} {101}},\ \bibinfo
  {pages} {103515} (\bibinfo {year} {2020})},\ \Eprint
  {https://arxiv.org/abs/2001.05909} {arXiv:2001.05909 [gr-qc]} \BibitemShut
  {NoStop}%
\bibitem [{\citenamefont {Yi}\ {\it et~al.}(2020)\citenamefont {Yi},
  \citenamefont {Gong}, \citenamefont {Wang},\ and\ \citenamefont
  {Zhu}}]{Yi:2020kmq}%
  \BibitemOpen
  \bibfield  {author} {\bibinfo {author} {\bibfnamefont {Z.}~\bibnamefont
  {Yi}}, \bibinfo {author} {\bibfnamefont {Y.}~\bibnamefont {Gong}}, \bibinfo
  {author} {\bibfnamefont {B.}~\bibnamefont {Wang}},\ and\ \bibinfo {author}
  {\bibfnamefont {Z.-H.}\ \bibnamefont {Zhu}},\ }\bibinfo {title} {{Primordial
  Black Holes and Secondary Gravitational Waves from Higgs field}},\ \href@noop
  {}  \Eprint {https://arxiv.org/abs/2007.09957}
  {arXiv:2007.09957 [gr-qc]} \BibitemShut {NoStop}%
\bibitem [{\citenamefont {Brans}\ and\ \citenamefont
  {Dicke}(1961)}]{Brans:1961sx}%
  \BibitemOpen
  \bibfield  {author} {\bibinfo {author} {\bibfnamefont {C.}~\bibnamefont
  {Brans}}\ and\ \bibinfo {author} {\bibfnamefont {R.~H.}\ \bibnamefont
  {Dicke}},\ }\bibinfo {title} {{Mach's principle and a relativistic theory of
  gravitation}},\ \href {https://doi.org/10.1103/PhysRev.124.925} {\bibfield
  {journal} {\bibinfo  {journal} {Phys. Rev.}\ }\textbf {\bibinfo {volume}
  {124}},\ \bibinfo {pages} {925} (\bibinfo {year} {1961})}\BibitemShut
  {NoStop}%
\bibitem [{\citenamefont {Dicke}(1962)}]{Dicke:1961gz}%
  \BibitemOpen
  \bibfield  {author} {\bibinfo {author} {\bibfnamefont {R.~H.}\ \bibnamefont
  {Dicke}},\ }\bibinfo {title} {{Mach's principle and invariance under
  transformation of units}},\ \href {https://doi.org/10.1103/PhysRev.125.2163}
  {\bibfield  {journal} {\bibinfo  {journal} {Phys. Rev.}\ }\textbf {\bibinfo
  {volume} {125}},\ \bibinfo {pages} {2163} (\bibinfo {year}
  {1962})}\BibitemShut {NoStop}%
\bibitem [{\citenamefont {Deffayet}\ {\it et~al.}(2011)\citenamefont
  {Deffayet}, \citenamefont {Gao}, \citenamefont {Steer},\ and\ \citenamefont
  {Zahariade}}]{Deffayet:2011gz}%
  \BibitemOpen
  \bibfield  {author} {\bibinfo {author} {\bibfnamefont {C.}~\bibnamefont
  {Deffayet}}, \bibinfo {author} {\bibfnamefont {X.}~\bibnamefont {Gao}},
  \bibinfo {author} {\bibfnamefont {D.~A.}\ \bibnamefont {Steer}},\ and\
  \bibinfo {author} {\bibfnamefont {G.}~\bibnamefont {Zahariade}},\ }\bibinfo
  {title} {{From k-essence to generalised Galileons}},\ \href
  {https://doi.org/10.1103/PhysRevD.84.064039} {\bibfield  {journal} {\bibinfo
  {journal} {Phys. Rev. D}\ }\textbf {\bibinfo {volume} {84}},\ \bibinfo
  {pages} {064039} (\bibinfo {year} {2011})},\ \Eprint
  {https://arxiv.org/abs/1103.3260} {arXiv:1103.3260 [hep-th]} \BibitemShut
  {NoStop}%
\bibitem [{\citenamefont {Kobayashi}\ {\it et~al.}(2011)\citenamefont
  {Kobayashi}, \citenamefont {Yamaguchi},\ and\ \citenamefont
  {Yokoyama}}]{Kobayashi:2011nu}%
  \BibitemOpen
  \bibfield  {author} {\bibinfo {author} {\bibfnamefont {T.}~\bibnamefont
  {Kobayashi}}, \bibinfo {author} {\bibfnamefont {M.}~\bibnamefont
  {Yamaguchi}},\ and\ \bibinfo {author} {\bibfnamefont {J.}~\bibnamefont
  {Yokoyama}},\ }\bibinfo {title} {{Generalized G-inflation: Inflation with the
  most general second-order field equations}},\ \href
  {https://doi.org/10.1143/PTP.126.511} {\bibfield  {journal} {\bibinfo
  {journal} {Prog. Theor. Phys.}\ }\textbf {\bibinfo {volume} {126}},\ \bibinfo
  {pages} {511} (\bibinfo {year} {2011})},\ \Eprint
  {https://arxiv.org/abs/1105.5723} {arXiv:1105.5723 [hep-th]} \BibitemShut
  {NoStop}%
\bibitem [{\citenamefont {Deffayet}\ {\it et~al.}(2009)\citenamefont
  {Deffayet}, \citenamefont {Esposito-Farese},\ and\ \citenamefont
  {Vikman}}]{Deffayet:2009wt}%
  \BibitemOpen
  \bibfield  {author} {\bibinfo {author} {\bibfnamefont {C.}~\bibnamefont
  {Deffayet}}, \bibinfo {author} {\bibfnamefont {G.}~\bibnamefont
  {Esposito-Farese}},\ and\ \bibinfo {author} {\bibfnamefont {A.}~\bibnamefont
  {Vikman}},\ }\bibinfo {title} {{Covariant Galileon}},\ \href
  {https://doi.org/10.1103/PhysRevD.79.084003} {\bibfield  {journal} {\bibinfo
  {journal} {Phys. Rev. D}\ }\textbf {\bibinfo {volume} {79}},\ \bibinfo
  {pages} {084003} (\bibinfo {year} {2009})},\ \Eprint
  {https://arxiv.org/abs/0901.1314} {arXiv:0901.1314 [hep-th]} \BibitemShut
  {NoStop}%
\bibitem [{\citenamefont {Nicolis}\ {\it et~al.}(2009)\citenamefont {Nicolis},
  \citenamefont {Rattazzi},\ and\ \citenamefont
  {Trincherini}}]{Nicolis:2008in}%
  \BibitemOpen
  \bibfield  {author} {\bibinfo {author} {\bibfnamefont {A.}~\bibnamefont
  {Nicolis}}, \bibinfo {author} {\bibfnamefont {R.}~\bibnamefont {Rattazzi}},\
  and\ \bibinfo {author} {\bibfnamefont {E.}~\bibnamefont {Trincherini}},\
  }\bibinfo {title} {{The Galileon as a local modification of gravity}},\ \href
  {https://doi.org/10.1103/PhysRevD.79.064036} {\bibfield  {journal} {\bibinfo
  {journal} {Phys. Rev. D}\ }\textbf {\bibinfo {volume} {79}},\ \bibinfo
  {pages} {064036} (\bibinfo {year} {2009})},\ \Eprint
  {https://arxiv.org/abs/0811.2197} {arXiv:0811.2197 [hep-th]} \BibitemShut
  {NoStop}%
\bibitem [{\citenamefont {Banerjee}\ and\ \citenamefont
  {Mukherjee}(2017)}]{Banerjee:2017qdl}%
  \BibitemOpen
  \bibfield  {author} {\bibinfo {author} {\bibfnamefont {R.}~\bibnamefont
  {Banerjee}}\ and\ \bibinfo {author} {\bibfnamefont {P.}~\bibnamefont
  {Mukherjee}},\ }\bibinfo {title} {{Taming galileons in curved spacetime}},\
  \href {https://doi.org/10.1088/1361-6382/aa9355} {\bibfield  {journal}
  {\bibinfo  {journal} {Class. Quant. Grav.}\ }\textbf {\bibinfo {volume}
  {34}},\ \bibinfo {pages} {235005} (\bibinfo {year} {2017})},\ \Eprint
  {https://arxiv.org/abs/1701.00971} {arXiv:1701.00971 [gr-qc]} \BibitemShut
  {NoStop}%
\bibitem [{\citenamefont {Bekenstein}(1993)}]{Bekenstein:1992pj}%
  \BibitemOpen
  \bibfield  {author} {\bibinfo {author} {\bibfnamefont {J.~D.}\ \bibnamefont
  {Bekenstein}},\ }\bibinfo {title} {{The Relation between physical and
  gravitational geometry}},\ \href {https://doi.org/10.1103/PhysRevD.48.3641}
  {\bibfield  {journal} {\bibinfo  {journal} {Phys. Rev. D}\ }\textbf {\bibinfo
  {volume} {48}},\ \bibinfo {pages} {3641} (\bibinfo {year} {1993})},\ \Eprint
  {https://arxiv.org/abs/gr-qc/9211017} {arXiv:gr-qc/9211017} \BibitemShut
  {NoStop}%
\bibitem [{\citenamefont {Zumalac\'arregui}\ and\ \citenamefont
  {Garc\'\i{}a-Bellido}(2014)}]{Zumalacarregui:2013pma}%
  \BibitemOpen
  \bibfield  {author} {\bibinfo {author} {\bibfnamefont {M.}~\bibnamefont
  {Zumalac\'arregui}}\ and\ \bibinfo {author} {\bibfnamefont {J.}~\bibnamefont
  {Garc\'\i{}a-Bellido}},\ }\bibinfo {title} {{Transforming gravity: from
  derivative couplings to matter to second-order scalar-tensor theories beyond
  the Horndeski Lagrangian}},\ \href
  {https://doi.org/10.1103/PhysRevD.89.064046} {\bibfield  {journal} {\bibinfo
  {journal} {Phys. Rev. D}\ }\textbf {\bibinfo {volume} {89}},\ \bibinfo
  {pages} {064046} (\bibinfo {year} {2014})},\ \Eprint
  {https://arxiv.org/abs/1308.4685} {arXiv:1308.4685 [gr-qc]} \BibitemShut
  {NoStop}%
\bibitem [{\citenamefont {Bettoni}\ and\ \citenamefont
  {Liberati}(2013)}]{Bettoni:2013diz}%
  \BibitemOpen
  \bibfield  {author} {\bibinfo {author} {\bibfnamefont {D.}~\bibnamefont
  {Bettoni}}\ and\ \bibinfo {author} {\bibfnamefont {S.}~\bibnamefont
  {Liberati}},\ }\bibinfo {title} {{Disformal invariance of second order
  scalar-tensor theories: Framing the Horndeski action}},\ \href
  {https://doi.org/10.1103/PhysRevD.88.084020} {\bibfield  {journal} {\bibinfo
  {journal} {Phys. Rev. D}\ }\textbf {\bibinfo {volume} {88}},\ \bibinfo
  {pages} {084020} (\bibinfo {year} {2013})},\ \Eprint
  {https://arxiv.org/abs/1306.6724} {arXiv:1306.6724 [gr-qc]} \BibitemShut
  {NoStop}%
\bibitem [{\citenamefont {Langlois}\ and\ \citenamefont
  {Noui}(2016{\natexlab{a}})}]{Langlois:2015cwa}%
  \BibitemOpen
  \bibfield  {author} {\bibinfo {author} {\bibfnamefont {D.}~\bibnamefont
  {Langlois}}\ and\ \bibinfo {author} {\bibfnamefont {K.}~\bibnamefont
  {Noui}},\ }\bibinfo {title} {{Degenerate higher derivative theories beyond
  Horndeski: evading the Ostrogradski instability}},\ \href
  {https://doi.org/10.1088/1475-7516/2016/02/034} {J. Cosmol. Astropart. Phys.\
  \bibinfo {volume} {02}\bibfield  {year} {\bibinfo  {year} { (\textbf
  {2016})}\ }\bibfield  {pages} {\bibinfo  {pages} {034}},\ }\Eprint
  {https://arxiv.org/abs/1510.06930} {arXiv:1510.06930 [gr-qc]} \BibitemShut
  {NoStop}%
\bibitem [{\citenamefont {Arroja}\ {\it et~al.}(2015)\citenamefont {Arroja},
  \citenamefont {Bartolo}, \citenamefont {Karmakar},\ and\ \citenamefont
  {Matarrese}}]{Arroja:2015wpa}%
  \BibitemOpen
  \bibfield  {author} {\bibinfo {author} {\bibfnamefont {F.}~\bibnamefont
  {Arroja}}, \bibinfo {author} {\bibfnamefont {N.}~\bibnamefont {Bartolo}},
  \bibinfo {author} {\bibfnamefont {P.}~\bibnamefont {Karmakar}},\ and\
  \bibinfo {author} {\bibfnamefont {S.}~\bibnamefont {Matarrese}},\ }\bibinfo
  {title} {{The two faces of mimetic Horndeski gravity: disformal
  transformations and Lagrange multiplier}},\ \href
  {https://doi.org/10.1088/1475-7516/2015/09/051} {J. Cosmol. Astropart. Phys.\
  \bibinfo {volume} {09}\bibfield  {year} {\bibinfo  {year} { (\textbf
  {2015})}\ }\bibfield  {pages} {\bibinfo  {pages} {051}},\ }\Eprint
  {https://arxiv.org/abs/1506.08575} {arXiv:1506.08575 [gr-qc]} \BibitemShut
  {NoStop}%
\bibitem [{\citenamefont {Langlois}\ and\ \citenamefont
  {Noui}(2016{\natexlab{b}})}]{Langlois:2015skt}%
  \BibitemOpen
  \bibfield  {author} {\bibinfo {author} {\bibfnamefont {D.}~\bibnamefont
  {Langlois}}\ and\ \bibinfo {author} {\bibfnamefont {K.}~\bibnamefont
  {Noui}},\ }\bibinfo {title} {{Hamiltonian analysis of higher derivative
  scalar-tensor theories}},\ \href
  {https://doi.org/10.1088/1475-7516/2016/07/016} {J. Cosmol. Astropart. Phys.\
  \bibinfo {volume} {07}\bibfield  {year} {\bibinfo  {year} { (\textbf
  {2016})}\ }\bibfield  {pages} {\bibinfo  {pages} {016}},\ }\Eprint
  {https://arxiv.org/abs/1512.06820} {arXiv:1512.06820 [gr-qc]} \BibitemShut
  {NoStop}%
\bibitem [{\citenamefont {Crisostomi}\ {\it
  et~al.}(2016{\natexlab{a}})\citenamefont {Crisostomi}, \citenamefont
  {Koyama},\ and\ \citenamefont {Tasinato}}]{Crisostomi:2016czh}%
  \BibitemOpen
  \bibfield  {author} {\bibinfo {author} {\bibfnamefont {M.}~\bibnamefont
  {Crisostomi}}, \bibinfo {author} {\bibfnamefont {K.}~\bibnamefont {Koyama}},\
  and\ \bibinfo {author} {\bibfnamefont {G.}~\bibnamefont {Tasinato}},\
  }\bibinfo {title} {{Extended Scalar-Tensor Theories of Gravity}},\ \href
  {https://doi.org/10.1088/1475-7516/2016/04/044} {J. Cosmol. Astropart. Phys.\
  \bibinfo {volume} {04}\bibfield  {year} {\bibinfo  {year} { (\textbf
  {2016})}\ }\bibfield  {pages} {\bibinfo  {pages} {044}},\ }\Eprint
  {https://arxiv.org/abs/1602.03119} {arXiv:1602.03119 [hep-th]} \BibitemShut
  {NoStop}%
\bibitem [{\citenamefont {Ben~Achour}\ {\it
  et~al.}(2016{\natexlab{a}})\citenamefont {Ben~Achour}, \citenamefont
  {Langlois},\ and\ \citenamefont {Noui}}]{Achour:2016rkg}%
  \BibitemOpen
  \bibfield  {author} {\bibinfo {author} {\bibfnamefont {J.}~\bibnamefont
  {Ben~Achour}}, \bibinfo {author} {\bibfnamefont {D.}~\bibnamefont
  {Langlois}},\ and\ \bibinfo {author} {\bibfnamefont {K.}~\bibnamefont
  {Noui}},\ }\bibinfo {title} {{Degenerate higher order scalar-tensor theories
  beyond Horndeski and disformal transformations}},\ \href
  {https://doi.org/10.1103/PhysRevD.93.124005} {\bibfield  {journal} {\bibinfo
  {journal} {Phys. Rev. D}\ }\textbf {\bibinfo {volume} {93}},\ \bibinfo
  {pages} {124005} (\bibinfo {year} {2016}{\natexlab{a}})},\ \Eprint
  {https://arxiv.org/abs/1602.08398} {arXiv:1602.08398 [gr-qc]} \BibitemShut
  {NoStop}%
\bibitem [{\citenamefont {Crisostomi}\ {\it
  et~al.}(2016{\natexlab{b}})\citenamefont {Crisostomi}, \citenamefont {Hull},
  \citenamefont {Koyama},\ and\ \citenamefont {Tasinato}}]{Crisostomi:2016tcp}%
  \BibitemOpen
  \bibfield  {author} {\bibinfo {author} {\bibfnamefont {M.}~\bibnamefont
  {Crisostomi}}, \bibinfo {author} {\bibfnamefont {M.}~\bibnamefont {Hull}},
  \bibinfo {author} {\bibfnamefont {K.}~\bibnamefont {Koyama}},\ and\ \bibinfo
  {author} {\bibfnamefont {G.}~\bibnamefont {Tasinato}},\ }\bibinfo {title}
  {{Horndeski: beyond, or not beyond?}},\ \href
  {https://doi.org/10.1088/1475-7516/2016/03/038} {J. Cosmol. Astropart. Phys.\
  \bibinfo {volume} {03}\bibfield  {year} {\bibinfo  {year} { (\textbf
  {2016})}\ }\bibfield  {pages} {\bibinfo  {pages} {038}},\ }\Eprint
  {https://arxiv.org/abs/1601.04658} {arXiv:1601.04658 [hep-th]} \BibitemShut
  {NoStop}%
\bibitem [{\citenamefont {Ben~Achour}\ {\it
  et~al.}(2016{\natexlab{b}})\citenamefont {Ben~Achour}, \citenamefont
  {Crisostomi}, \citenamefont {Koyama}, \citenamefont {Langlois}, \citenamefont
  {Noui},\ and\ \citenamefont {Tasinato}}]{BenAchour:2016fzp}%
  \BibitemOpen
  \bibfield  {author} {\bibinfo {author} {\bibfnamefont {J.}~\bibnamefont
  {Ben~Achour}}, \bibinfo {author} {\bibfnamefont {M.}~\bibnamefont
  {Crisostomi}}, \bibinfo {author} {\bibfnamefont {K.}~\bibnamefont {Koyama}},
  \bibinfo {author} {\bibfnamefont {D.}~\bibnamefont {Langlois}}, \bibinfo
  {author} {\bibfnamefont {K.}~\bibnamefont {Noui}},\ and\ \bibinfo {author}
  {\bibfnamefont {G.}~\bibnamefont {Tasinato}},\ }\bibinfo {title} {{Degenerate
  higher order scalar-tensor theories beyond Horndeski up to cubic order}},\
  \href {https://doi.org/10.1007/JHEP12(2016)100} {J. High Energ. Phys.\
  \bibinfo {volume} {12}\bibfield  {year} {\bibinfo  {year} { (\textbf
  {2016})}\ }\bibfield  {pages} {\bibinfo  {pages} {100}},\ }\Eprint
  {https://arxiv.org/abs/1608.08135} {arXiv:1608.08135 [hep-th]} \BibitemShut
  {NoStop}%
\bibitem [{\citenamefont {Langlois}\ {\it et~al.}(2017)\citenamefont
  {Langlois}, \citenamefont {Mancarella}, \citenamefont {Noui},\ and\
  \citenamefont {Vernizzi}}]{Langlois:2017mxy}%
  \BibitemOpen
  \bibfield  {author} {\bibinfo {author} {\bibfnamefont {D.}~\bibnamefont
  {Langlois}}, \bibinfo {author} {\bibfnamefont {M.}~\bibnamefont
  {Mancarella}}, \bibinfo {author} {\bibfnamefont {K.}~\bibnamefont {Noui}},\
  and\ \bibinfo {author} {\bibfnamefont {F.}~\bibnamefont {Vernizzi}},\
  }\bibinfo {title} {{Effective Description of Higher-Order Scalar-Tensor
  Theories}},\ \href {https://doi.org/10.1088/1475-7516/2017/05/033} {J.
  Cosmol. Astropart. Phys.\ \bibinfo {volume} {05}\bibfield  {year} {\bibinfo
  {year} { (\textbf {2017})}\ }\bibfield  {pages} {\bibinfo  {pages} {033}},\
  }\Eprint {https://arxiv.org/abs/1703.03797} {arXiv:1703.03797 [hep-th]}
  \BibitemShut {NoStop}%
\bibitem [{\citenamefont {Takahashi}\ and\ \citenamefont
  {Kobayashi}(2017)}]{Takahashi:2017pje}%
  \BibitemOpen
  \bibfield  {author} {\bibinfo {author} {\bibfnamefont {K.}~\bibnamefont
  {Takahashi}}\ and\ \bibinfo {author} {\bibfnamefont {T.}~\bibnamefont
  {Kobayashi}},\ }\bibinfo {title} {{Extended mimetic gravity: Hamiltonian
  analysis and gradient instabilities}},\ \href
  {https://doi.org/10.1088/1475-7516/2017/11/038} {J. Cosmol. Astropart. Phys.\
  \bibinfo {volume} {11}\bibfield  {year} {\bibinfo  {year} { (\textbf
  {2017})}\ }\bibfield  {pages} {\bibinfo  {pages} {038}},\ }\Eprint
  {https://arxiv.org/abs/1708.02951} {arXiv:1708.02951 [gr-qc]} \BibitemShut
  {NoStop}%
\bibitem [{\citenamefont {Motohashi}\ {\it et~al.}(2016)\citenamefont
  {Motohashi}, \citenamefont {Noui}, \citenamefont {Suyama}, \citenamefont
  {Yamaguchi},\ and\ \citenamefont {Langlois}}]{Motohashi:2016ftl}%
  \BibitemOpen
  \bibfield  {author} {\bibinfo {author} {\bibfnamefont {H.}~\bibnamefont
  {Motohashi}}, \bibinfo {author} {\bibfnamefont {K.}~\bibnamefont {Noui}},
  \bibinfo {author} {\bibfnamefont {T.}~\bibnamefont {Suyama}}, \bibinfo
  {author} {\bibfnamefont {M.}~\bibnamefont {Yamaguchi}},\ and\ \bibinfo
  {author} {\bibfnamefont {D.}~\bibnamefont {Langlois}},\ }\bibinfo {title}
  {{Healthy degenerate theories with higher derivatives}},\ \href
  {https://doi.org/10.1088/1475-7516/2016/07/033} {J. Cosmol. Astropart. Phys.\
  \bibinfo {volume} {07}\bibfield  {year} {\bibinfo  {year} { (\textbf
  {2016})}\ }\bibfield  {pages} {\bibinfo  {pages} {033}},\ }\Eprint
  {https://arxiv.org/abs/1603.09355} {arXiv:1603.09355 [hep-th]} \BibitemShut
  {NoStop}%
\bibitem [{\citenamefont {Klein}\ and\ \citenamefont
  {Roest}(2016)}]{Klein:2016aiq}%
  \BibitemOpen
  \bibfield  {author} {\bibinfo {author} {\bibfnamefont {R.}~\bibnamefont
  {Klein}}\ and\ \bibinfo {author} {\bibfnamefont {D.}~\bibnamefont {Roest}},\
  }\bibinfo {title} {{Exorcising the Ostrogradsky ghost in coupled systems}},\
  \href {https://doi.org/10.1007/JHEP07(2016)130} {J. High Energ. Phys.\
  \bibinfo {volume} {07}\bibfield  {year} {\bibinfo  {year} { (\textbf
  {2016})}\ }\bibfield  {pages} {\bibinfo  {pages} {130}},\ }\Eprint
  {https://arxiv.org/abs/1604.01719} {arXiv:1604.01719 [hep-th]} \BibitemShut
  {NoStop}%
\bibitem [{\citenamefont {Ostrogradsky}(1850)}]{Ostrogradsky:1850fid}%
  \BibitemOpen
  \bibfield  {author} {\bibinfo {author} {\bibfnamefont {M.}~\bibnamefont
  {Ostrogradsky}},\ }\bibinfo {title} {{M\'emoires sur les \'equations
  diff\'erentielles, relatives au probl\`eme des isop\'erim\`etres}},\
  \href@noop {} {\bibfield  {journal} {\bibinfo  {journal} {Mem. Acad. St.
  Petersbourg}\ }\textbf {\bibinfo {volume} {6}},\ \bibinfo {pages} {385}
  (\bibinfo {year} {1850})}\BibitemShut {NoStop}%
\bibitem [{\citenamefont {Woodard}(2015)}]{Woodard:2015zca}%
  \BibitemOpen
  \bibfield  {author} {\bibinfo {author} {\bibfnamefont {R.~P.}\ \bibnamefont
  {Woodard}},\ }\bibinfo {title} {{Ostrogradsky's theorem on Hamiltonian
  instability}},\ \href {https://doi.org/10.4249/scholarpedia.32243} {\bibfield
   {journal} {\bibinfo  {journal} {Scholarpedia}\ }\textbf {\bibinfo {volume}
  {10}},\ \bibinfo {pages} {32243} (\bibinfo {year} {2015})},\ \Eprint
  {https://arxiv.org/abs/1506.02210} {arXiv:1506.02210 [hep-th]} \BibitemShut
  {NoStop}%
\bibitem [{\citenamefont {Khoury}\ {\it et~al.}(2012)\citenamefont {Khoury},
  \citenamefont {Miller},\ and\ \citenamefont {Tolley}}]{Khoury:2011ay}%
  \BibitemOpen
  \bibfield  {author} {\bibinfo {author} {\bibfnamefont {J.}~\bibnamefont
  {Khoury}}, \bibinfo {author} {\bibfnamefont {G.~E.~J.}\ \bibnamefont
  {Miller}},\ and\ \bibinfo {author} {\bibfnamefont {A.~J.}\ \bibnamefont
  {Tolley}},\ }\bibinfo {title} {{Spatially Covariant Theories of a Transverse,
  Traceless Graviton, Part I: Formalism}},\ \href
  {https://doi.org/10.1103/PhysRevD.85.084002} {\bibfield  {journal} {\bibinfo
  {journal} {Phys. Rev. D}\ }\textbf {\bibinfo {volume} {85}},\ \bibinfo
  {pages} {084002} (\bibinfo {year} {2012})},\ \Eprint
  {https://arxiv.org/abs/1108.1397} {arXiv:1108.1397 [hep-th]} \BibitemShut
  {NoStop}%
\bibitem [{\citenamefont {Fujita}\ {\it et~al.}(2016)\citenamefont {Fujita},
  \citenamefont {Gao},\ and\ \citenamefont {Yokoyama}}]{Fujita:2015ymn}%
  \BibitemOpen
  \bibfield  {author} {\bibinfo {author} {\bibfnamefont {T.}~\bibnamefont
  {Fujita}}, \bibinfo {author} {\bibfnamefont {X.}~\bibnamefont {Gao}},\ and\
  \bibinfo {author} {\bibfnamefont {J.}~\bibnamefont {Yokoyama}},\ }\bibinfo
  {title} {{Spatially covariant theories of gravity: disformal transformation,
  cosmological perturbations and the Einstein frame}},\ \href
  {https://doi.org/10.1088/1475-7516/2016/02/014} {J. Cosmol. Astropart. Phys.\
  \bibinfo {volume} {02}\bibfield  {year} {\bibinfo  {year} { (\textbf
  {2016})}\ }\bibfield  {pages} {\bibinfo  {pages} {014}},\ }\Eprint
  {https://arxiv.org/abs/1511.04324} {arXiv:1511.04324 [gr-qc]} \BibitemShut
  {NoStop}%
\bibitem [{\citenamefont {Gao}(2014{\natexlab{a}})}]{Gao:2014soa}%
  \BibitemOpen
  \bibfield  {author} {\bibinfo {author} {\bibfnamefont {X.}~\bibnamefont
  {Gao}},\ }\bibinfo {title} {{Unifying framework for scalar-tensor theories of
  gravity}},\ \href {https://doi.org/10.1103/PhysRevD.90.081501} {\bibfield
  {journal} {\bibinfo  {journal} {Phys. Rev. D}\ }\textbf {\bibinfo {volume}
  {90}},\ \bibinfo {pages} {081501} (\bibinfo {year} {2014}{\natexlab{a}})},\
  \Eprint {https://arxiv.org/abs/1406.0822} {arXiv:1406.0822 [gr-qc]}
  \BibitemShut {NoStop}%
\bibitem [{\citenamefont {Gao}(2014{\natexlab{b}})}]{Gao:2014fra}%
  \BibitemOpen
  \bibfield  {author} {\bibinfo {author} {\bibfnamefont {X.}~\bibnamefont
  {Gao}},\ }\bibinfo {title} {{Hamiltonian analysis of spatially covariant
  gravity}},\ \href {https://doi.org/10.1103/PhysRevD.90.104033} {\bibfield
  {journal} {\bibinfo  {journal} {Phys. Rev. D}\ }\textbf {\bibinfo {volume}
  {90}},\ \bibinfo {pages} {104033} (\bibinfo {year} {2014}{\natexlab{b}})},\
  \Eprint {https://arxiv.org/abs/1409.6708} {arXiv:1409.6708 [gr-qc]}
  \BibitemShut {NoStop}%
\bibitem [{\citenamefont {Gao}\ and\ \citenamefont {Yao}(2019)}]{Gao:2018znj}%
  \BibitemOpen
  \bibfield  {author} {\bibinfo {author} {\bibfnamefont {X.}~\bibnamefont
  {Gao}}\ and\ \bibinfo {author} {\bibfnamefont {Z.-B.}\ \bibnamefont {Yao}},\
  }\bibinfo {title} {{Spatially covariant gravity with velocity of the lapse
  function: the Hamiltonian analysis}},\ \href
  {https://doi.org/10.1088/1475-7516/2019/05/024} {J. Cosmol. Astropart. Phys.\
  \bibinfo {volume} {05}\bibfield  {year} {\bibinfo  {year} { (\textbf
  {2019})}\ }\bibfield  {pages} {\bibinfo  {pages} {024}},\ }\Eprint
  {https://arxiv.org/abs/1806.02811} {arXiv:1806.02811 [gr-qc]} \BibitemShut
  {NoStop}%
\bibitem [{\citenamefont {Gao}\ {\it et~al.}(2019{\natexlab{a}})\citenamefont
  {Gao}, \citenamefont {Yamaguchi},\ and\ \citenamefont
  {Yoshida}}]{Gao:2018izs}%
  \BibitemOpen
  \bibfield  {author} {\bibinfo {author} {\bibfnamefont {X.}~\bibnamefont
  {Gao}}, \bibinfo {author} {\bibfnamefont {M.}~\bibnamefont {Yamaguchi}},\
  and\ \bibinfo {author} {\bibfnamefont {D.}~\bibnamefont {Yoshida}},\
  }\bibinfo {title} {{Higher derivative scalar-tensor theory through a
  non-dynamical scalar field}},\ \href
  {https://doi.org/10.1088/1475-7516/2019/03/006} {J. Cosmol. Astropart. Phys.\
  \bibinfo {volume} {03}\bibfield  {year} {\bibinfo  {year} { (\textbf
  {2019})}\ }\bibfield  {pages} {\bibinfo  {pages} {006}},\ }\Eprint
  {https://arxiv.org/abs/1810.07434} {arXiv:1810.07434 [hep-th]} \BibitemShut
  {NoStop}%
\bibitem [{\citenamefont {Gao}\ {\it et~al.}(2019{\natexlab{b}})\citenamefont
  {Gao}, \citenamefont {Kang},\ and\ \citenamefont {Yao}}]{Gao:2019lpz}%
  \BibitemOpen
  \bibfield  {author} {\bibinfo {author} {\bibfnamefont {X.}~\bibnamefont
  {Gao}}, \bibinfo {author} {\bibfnamefont {C.}~\bibnamefont {Kang}},\ and\
  \bibinfo {author} {\bibfnamefont {Z.-B.}\ \bibnamefont {Yao}},\ }\bibinfo
  {title} {{Spatially Covariant Gravity: Perturbative Analysis and Field
  Transformations}},\ \href {https://doi.org/10.1103/PhysRevD.99.104015}
  {\bibfield  {journal} {\bibinfo  {journal} {Phys. Rev. D}\ }\textbf {\bibinfo
  {volume} {99}},\ \bibinfo {pages} {104015} (\bibinfo {year}
  {2019}{\natexlab{b}})},\ \Eprint {https://arxiv.org/abs/1902.07702}
  {arXiv:1902.07702 [gr-qc]} \BibitemShut {NoStop}%
\bibitem [{\citenamefont {Gao}\ and\ \citenamefont {Yao}(2020)}]{Gao:2019twq}%
  \BibitemOpen
  \bibfield  {author} {\bibinfo {author} {\bibfnamefont {X.}~\bibnamefont
  {Gao}}\ and\ \bibinfo {author} {\bibfnamefont {Z.-B.}\ \bibnamefont {Yao}},\
  }\bibinfo {title} {{Spatially covariant gravity theories with two tensorial
  degrees of freedom: the formalism}},\ \href
  {https://doi.org/10.1103/PhysRevD.101.064018} {\bibfield  {journal} {\bibinfo
   {journal} {Phys. Rev. D}\ }\textbf {\bibinfo {volume} {101}},\ \bibinfo
  {pages} {064018} (\bibinfo {year} {2020})},\ \Eprint
  {https://arxiv.org/abs/1910.13995} {arXiv:1910.13995 [gr-qc]} \BibitemShut
  {NoStop}%
\bibitem [{\citenamefont {Gao}(2021)}]{Gao:2020juc}%
  \BibitemOpen
  \bibfield  {author} {\bibinfo {author} {\bibfnamefont {X.}~\bibnamefont
  {Gao}},\ }\bibinfo {title} {{Higher derivative scalar-tensor monomials and
  their classification}},\ \href {https://doi.org/10.1007/s11433-020-1607-3}
  {\bibfield  {journal} {\bibinfo  {journal} {Sci. China Phys. Mech. Astron.}\
  }\textbf {\bibinfo {volume} {64}},\ \bibinfo {pages} {210012} (\bibinfo
  {year} {2021})},\ \Eprint {https://arxiv.org/abs/2003.11978}
  {arXiv:2003.11978 [gr-qc]} \BibitemShut {NoStop}%
\bibitem [{\citenamefont {Gao}\ and\ \citenamefont {Hu}(2020)}]{Gao:2020yzr}%
  \BibitemOpen
  \bibfield  {author} {\bibinfo {author} {\bibfnamefont {X.}~\bibnamefont
  {Gao}}\ and\ \bibinfo {author} {\bibfnamefont {Y.-M.}\ \bibnamefont {Hu}},\
  }\bibinfo {title} {{Higher derivative scalar-tensor theory and spatially
  covariant gravity: the correspondence}},\ \href
  {https://doi.org/10.1103/PhysRevD.102.084006} {\bibfield  {journal} {\bibinfo
   {journal} {Phys. Rev. D}\ }\textbf {\bibinfo {volume} {102}},\ \bibinfo
  {pages} {084006} (\bibinfo {year} {2020})},\ \Eprint
  {https://arxiv.org/abs/2004.07752} {arXiv:2004.07752 [gr-qc]} \BibitemShut
  {NoStop}%
\bibitem [{\citenamefont {Gao}(2020)}]{Gao:2020qxy}%
  \BibitemOpen
  \bibfield  {author} {\bibinfo {author} {\bibfnamefont {X.}~\bibnamefont
  {Gao}},\ }\bibinfo {title} {{Higher derivative scalar-tensor theory from the
  spatially covariant gravity: a linear algebraic analysis}},\ \href
  {https://doi.org/10.1088/1475-7516/2020/11/004} {J. Cosmol. Astropart. Phys.\
  \bibinfo {volume} {11}\bibfield  {year} {\bibinfo  {year} { (\textbf
  {2020})}\ }\bibfield  {pages} {\bibinfo  {pages} {004}},\ }\Eprint
  {https://arxiv.org/abs/2006.15633} {arXiv:2006.15633 [gr-qc]} \BibitemShut
  {NoStop}%
\bibitem [{\citenamefont {Son}\ and\ \citenamefont
  {Wingate}(2006)}]{Son:2005rv}%
  \BibitemOpen
  \bibfield  {author} {\bibinfo {author} {\bibfnamefont {D.~T.}\ \bibnamefont
  {Son}}\ and\ \bibinfo {author} {\bibfnamefont {M.}~\bibnamefont {Wingate}},\
  }\bibinfo {title} {{General coordinate invariance and conformal invariance in
  nonrelativistic physics: Unitary Fermi gas}},\ \href
  {https://doi.org/10.1016/j.aop.2005.11.001} {\bibfield  {journal} {\bibinfo
  {journal} {Annals Phys.}\ }\textbf {\bibinfo {volume} {321}},\ \bibinfo
  {pages} {197} (\bibinfo {year} {2006})},\ \Eprint
  {https://arxiv.org/abs/cond-mat/0509786} {arXiv:cond-mat/0509786}
  \BibitemShut {NoStop}%
\bibitem [{\citenamefont {Banerjee}\ {\it et~al.}(2014)\citenamefont
  {Banerjee}, \citenamefont {Mitra},\ and\ \citenamefont
  {Mukherjee}}]{Banerjee:2014pya}%
  \BibitemOpen
  \bibfield  {author} {\bibinfo {author} {\bibfnamefont {R.}~\bibnamefont
  {Banerjee}}, \bibinfo {author} {\bibfnamefont {A.}~\bibnamefont {Mitra}},\
  and\ \bibinfo {author} {\bibfnamefont {P.}~\bibnamefont {Mukherjee}},\
  }\bibinfo {title} {{A new formulation of non-relativistic diffeomorphism
  invariance}},\ \href {https://doi.org/10.1016/j.physletb.2014.09.004}
  {\bibfield  {journal} {\bibinfo  {journal} {Phys. Lett. B}\ }\textbf
  {\bibinfo {volume} {737}},\ \bibinfo {pages} {369} (\bibinfo {year}
  {2014})},\ \Eprint {https://arxiv.org/abs/1404.4491} {arXiv:1404.4491
  [gr-qc]} \BibitemShut {NoStop}%
\bibitem [{\citenamefont {Banerjee}\ and\ \citenamefont
  {Mukherjee}(2016)}]{Banerjee:2015rca}%
  \BibitemOpen
  \bibfield  {author} {\bibinfo {author} {\bibfnamefont {R.}~\bibnamefont
  {Banerjee}}\ and\ \bibinfo {author} {\bibfnamefont {P.}~\bibnamefont
  {Mukherjee}},\ }\bibinfo {title} {{New approach to nonrelativistic
  diffeomorphism invariance and its applications}},\ \href
  {https://doi.org/10.1103/PhysRevD.93.085020} {\bibfield  {journal} {\bibinfo
  {journal} {Phys. Rev. D}\ }\textbf {\bibinfo {volume} {93}},\ \bibinfo
  {pages} {085020} (\bibinfo {year} {2016})},\ \Eprint
  {https://arxiv.org/abs/1509.05622} {arXiv:1509.05622 [gr-qc]} \BibitemShut
  {NoStop}%
\bibitem [{\citenamefont {Arkani-Hamed}\ {\it et~al.}(2004)\citenamefont
  {Arkani-Hamed}, \citenamefont {Cheng}, \citenamefont {Luty},\ and\
  \citenamefont {Mukohyama}}]{ArkaniHamed:2003uy}%
  \BibitemOpen
  \bibfield  {author} {\bibinfo {author} {\bibfnamefont {N.}~\bibnamefont
  {Arkani-Hamed}}, \bibinfo {author} {\bibfnamefont {H.-C.}\ \bibnamefont
  {Cheng}}, \bibinfo {author} {\bibfnamefont {M.~A.}\ \bibnamefont {Luty}},\
  and\ \bibinfo {author} {\bibfnamefont {S.}~\bibnamefont {Mukohyama}},\
  }\bibinfo {title} {{Ghost condensation and a consistent infrared modification
  of gravity}},\ \href {https://doi.org/10.1088/1126-6708/2004/05/074} {J. High
  Energ. Phys.\ \bibinfo {volume} {05}\bibfield  {year} {\bibinfo  {year} {
  (\textbf {2004})}\ }\bibfield  {pages} {\bibinfo  {pages} {074}},\ }\Eprint
  {https://arxiv.org/abs/hep-th/0312099} {arXiv:hep-th/0312099} \BibitemShut
  {NoStop}%
\bibitem [{\citenamefont {Creminelli}\ {\it et~al.}(2006)\citenamefont
  {Creminelli}, \citenamefont {Luty}, \citenamefont {Nicolis},\ and\
  \citenamefont {Senatore}}]{Creminelli:2006xe}%
  \BibitemOpen
  \bibfield  {author} {\bibinfo {author} {\bibfnamefont {P.}~\bibnamefont
  {Creminelli}}, \bibinfo {author} {\bibfnamefont {M.~A.}\ \bibnamefont
  {Luty}}, \bibinfo {author} {\bibfnamefont {A.}~\bibnamefont {Nicolis}},\ and\
  \bibinfo {author} {\bibfnamefont {L.}~\bibnamefont {Senatore}},\ }\bibinfo
  {title} {{Starting the Universe: Stable Violation of the Null Energy
  Condition and Non-standard Cosmologies}},\ \href
  {https://doi.org/10.1088/1126-6708/2006/12/080} {J. High Energ. Phys.\
  \bibinfo {volume} {12}\bibfield  {year} {\bibinfo  {year} { (\textbf
  {2006})}\ }\bibfield  {pages} {\bibinfo  {pages} {080}},\ }\Eprint
  {https://arxiv.org/abs/hep-th/0606090} {arXiv:hep-th/0606090} \BibitemShut
  {NoStop}%
\bibitem [{\citenamefont {Cheung}\ {\it et~al.}(2008)\citenamefont {Cheung},
  \citenamefont {Creminelli}, \citenamefont {Fitzpatrick}, \citenamefont
  {Kaplan},\ and\ \citenamefont {Senatore}}]{Cheung:2007st}%
  \BibitemOpen
  \bibfield  {author} {\bibinfo {author} {\bibfnamefont {C.}~\bibnamefont
  {Cheung}}, \bibinfo {author} {\bibfnamefont {P.}~\bibnamefont {Creminelli}},
  \bibinfo {author} {\bibfnamefont {A.~L.}\ \bibnamefont {Fitzpatrick}},
  \bibinfo {author} {\bibfnamefont {J.}~\bibnamefont {Kaplan}},\ and\ \bibinfo
  {author} {\bibfnamefont {L.}~\bibnamefont {Senatore}},\ }\bibinfo {title}
  {{The Effective Field Theory of Inflation}},\ \href
  {https://doi.org/10.1088/1126-6708/2008/03/014} {J. High Energ. Phys.\
  \bibinfo {volume} {03}\bibfield  {year} {\bibinfo  {year} { (\textbf
  {2008})}\ }\bibfield  {pages} {\bibinfo  {pages} {014}},\ }\Eprint
  {https://arxiv.org/abs/0709.0293} {arXiv:0709.0293 [hep-th]} \BibitemShut
  {NoStop}%
\bibitem [{\citenamefont {Horava}(2009{\natexlab{a}})}]{Horava:2009uw}%
  \BibitemOpen
  \bibfield  {author} {\bibinfo {author} {\bibfnamefont {P.}~\bibnamefont
  {Horava}},\ }\bibinfo {title} {{Quantum Gravity at a Lifshitz Point}},\ \href
  {https://doi.org/10.1103/PhysRevD.79.084008} {\bibfield  {journal} {\bibinfo
  {journal} {Phys. Rev. D}\ }\textbf {\bibinfo {volume} {79}},\ \bibinfo
  {pages} {084008} (\bibinfo {year} {2009}{\natexlab{a}})},\ \Eprint
  {https://arxiv.org/abs/0901.3775} {arXiv:0901.3775 [hep-th]} \BibitemShut
  {NoStop}%
\bibitem [{\citenamefont {Horava}(2009{\natexlab{b}})}]{Horava:2009if}%
  \BibitemOpen
  \bibfield  {author} {\bibinfo {author} {\bibfnamefont {P.}~\bibnamefont
  {Horava}},\ }\bibinfo {title} {{Spectral Dimension of the Universe in Quantum
  Gravity at a Lifshitz Point}},\ \href
  {https://doi.org/10.1103/PhysRevLett.102.161301} {\bibfield  {journal}
  {\bibinfo  {journal} {Phys. Rev. Lett.}\ }\textbf {\bibinfo {volume} {102}},\
  \bibinfo {pages} {161301} (\bibinfo {year} {2009}{\natexlab{b}})},\ \Eprint
  {https://arxiv.org/abs/0902.3657} {arXiv:0902.3657 [hep-th]} \BibitemShut
  {NoStop}%
\bibitem [{\citenamefont {Visser}(2009)}]{Visser:2009fg}%
  \BibitemOpen
  \bibfield  {author} {\bibinfo {author} {\bibfnamefont {M.}~\bibnamefont
  {Visser}},\ }\bibinfo {title} {{Lorentz symmetry breaking as a quantum field
  theory regulator}},\ \href {https://doi.org/10.1103/PhysRevD.80.025011}
  {\bibfield  {journal} {\bibinfo  {journal} {Phys. Rev. D}\ }\textbf {\bibinfo
  {volume} {80}},\ \bibinfo {pages} {025011} (\bibinfo {year} {2009})},\
  \Eprint {https://arxiv.org/abs/0902.0590} {arXiv:0902.0590 [hep-th]}
  \BibitemShut {NoStop}%
\bibitem [{\citenamefont {Devecioglu}\ and\ \citenamefont
  {Park}(2020)}]{Devecioglu:2020dny}%
  \BibitemOpen
  \bibfield  {author} {\bibinfo {author} {\bibfnamefont {D.~O.}\ \bibnamefont
  {Devecioglu}}\ and\ \bibinfo {author} {\bibfnamefont {M.-I.}\ \bibnamefont
  {Park}},\ }\bibinfo {title} {{The Hamiltonian dynamics of Ho\v{r}ava
  gravity}},\ \href {https://doi.org/10.1140/epjc/s10052-020-8139-8} {\bibfield
   {journal} {\bibinfo  {journal} {Eur. Phys. J. C}\ }\textbf {\bibinfo
  {volume} {80}},\ \bibinfo {pages} {597} (\bibinfo {year} {2020})},\ \bibinfo
  {note} {[Erratum: Eur. Phys. J. C 80, 764 (2020)]},\ \Eprint
  {https://arxiv.org/abs/2001.02556} {arXiv:2001.02556 [hep-th]} \BibitemShut
  {NoStop}%
\bibitem [{\citenamefont {Chamseddine}\ and\ \citenamefont
  {Mukhanov}(2013)}]{Chamseddine:2013kea}%
  \BibitemOpen
  \bibfield  {author} {\bibinfo {author} {\bibfnamefont {A.~H.}\ \bibnamefont
  {Chamseddine}}\ and\ \bibinfo {author} {\bibfnamefont {V.}~\bibnamefont
  {Mukhanov}},\ }\bibinfo {title} {{Mimetic Dark Matter}},\ \href
  {https://doi.org/10.1007/JHEP11(2013)135} {J. High Energ. Phys.\ \bibinfo
  {volume} {11}\bibfield  {year} {\bibinfo  {year} { (\textbf {2013})}\
  }\bibfield  {pages} {\bibinfo  {pages} {135}},\ }\Eprint
  {https://arxiv.org/abs/1308.5410} {arXiv:1308.5410 [astro-ph.CO]}
  \BibitemShut {NoStop}%
\bibitem [{\citenamefont {Sebastiani}\ {\it et~al.}(2017)\citenamefont
  {Sebastiani}, \citenamefont {Vagnozzi},\ and\ \citenamefont
  {Myrzakulov}}]{Sebastiani:2016ras}%
  \BibitemOpen
  \bibfield  {author} {\bibinfo {author} {\bibfnamefont {L.}~\bibnamefont
  {Sebastiani}}, \bibinfo {author} {\bibfnamefont {S.}~\bibnamefont
  {Vagnozzi}},\ and\ \bibinfo {author} {\bibfnamefont {R.}~\bibnamefont
  {Myrzakulov}},\ }\bibinfo {title} {{Mimetic gravity: a review of recent
  developments and applications to cosmology and astrophysics}},\ \href
  {https://doi.org/10.1155/2017/3156915} {\bibfield  {journal} {\bibinfo
  {journal} {Adv. High Energy Phys.}\ }\textbf {\bibinfo {volume} {2017}},\
  \bibinfo {pages} {3156915} (\bibinfo {year} {2017})},\ \Eprint
  {https://arxiv.org/abs/1612.08661} {arXiv:1612.08661 [gr-qc]} \BibitemShut
  {NoStop}%
\bibitem [{\citenamefont {Yao}\ {\it et~al.}(2021)\citenamefont {Yao},
  \citenamefont {Oliosi}, \citenamefont {Gao},\ and\ \citenamefont
  {Mukohyama}}]{Yao:2020tur}%
  \BibitemOpen
  \bibfield  {author} {\bibinfo {author} {\bibfnamefont {Z.-B.}\ \bibnamefont
  {Yao}}, \bibinfo {author} {\bibfnamefont {M.}~\bibnamefont {Oliosi}},
  \bibinfo {author} {\bibfnamefont {X.}~\bibnamefont {Gao}},\ and\ \bibinfo
  {author} {\bibfnamefont {S.}~\bibnamefont {Mukohyama}},\ }\bibinfo {title}
  {{Minimally modified gravity with an auxiliary constraint: a Hamiltonian
  construction}},\ \href {https://doi.org/10.1103/PhysRevD.103.024032}
  {\bibfield  {journal} {\bibinfo  {journal} {Phys. Rev. D}\ }\textbf {\bibinfo
  {volume} {103}},\ \bibinfo {pages} {024032} (\bibinfo {year} {2021})},\
  \Eprint {https://arxiv.org/abs/2011.00805} {arXiv:2011.00805 [gr-qc]}
  \BibitemShut {NoStop}%
\bibitem [{\citenamefont {Afshordi}\ {\it
  et~al.}(2007{\natexlab{a}})\citenamefont {Afshordi}, \citenamefont {Chung},\
  and\ \citenamefont {Geshnizjani}}]{Afshordi:2006ad}%
  \BibitemOpen
  \bibfield  {author} {\bibinfo {author} {\bibfnamefont {N.}~\bibnamefont
  {Afshordi}}, \bibinfo {author} {\bibfnamefont {D.~J.~H.}\ \bibnamefont
  {Chung}},\ and\ \bibinfo {author} {\bibfnamefont {G.}~\bibnamefont
  {Geshnizjani}},\ }\bibinfo {title} {{Cuscuton: A Causal Field Theory with an
  Infinite Speed of Sound}},\ \href
  {https://doi.org/10.1103/PhysRevD.75.083513} {\bibfield  {journal} {\bibinfo
  {journal} {Phys. Rev. D}\ }\textbf {\bibinfo {volume} {75}},\ \bibinfo
  {pages} {083513} (\bibinfo {year} {2007}{\natexlab{a}})},\ \Eprint
  {https://arxiv.org/abs/hep-th/0609150} {arXiv:hep-th/0609150} \BibitemShut
  {NoStop}%
\bibitem [{\citenamefont {Afshordi}\ {\it
  et~al.}(2007{\natexlab{b}})\citenamefont {Afshordi}, \citenamefont {Chung},
  \citenamefont {Doran},\ and\ \citenamefont {Geshnizjani}}]{Afshordi:2007yx}%
  \BibitemOpen
  \bibfield  {author} {\bibinfo {author} {\bibfnamefont {N.}~\bibnamefont
  {Afshordi}}, \bibinfo {author} {\bibfnamefont {D.~J.~H.}\ \bibnamefont
  {Chung}}, \bibinfo {author} {\bibfnamefont {M.}~\bibnamefont {Doran}},\ and\
  \bibinfo {author} {\bibfnamefont {G.}~\bibnamefont {Geshnizjani}},\ }\bibinfo
  {title} {{Cuscuton Cosmology: Dark Energy meets Modified Gravity}},\ \href
  {https://doi.org/10.1103/PhysRevD.75.123509} {\bibfield  {journal} {\bibinfo
  {journal} {Phys. Rev. D}\ }\textbf {\bibinfo {volume} {75}},\ \bibinfo
  {pages} {123509} (\bibinfo {year} {2007}{\natexlab{b}})},\ \Eprint
  {https://arxiv.org/abs/astro-ph/0702002} {arXiv:astro-ph/0702002}
  \BibitemShut {NoStop}%
\bibitem [{\citenamefont {Gomes}\ and\ \citenamefont
  {Guariento}(2017)}]{Gomes:2017tzd}%
  \BibitemOpen
  \bibfield  {author} {\bibinfo {author} {\bibfnamefont {H.}~\bibnamefont
  {Gomes}}\ and\ \bibinfo {author} {\bibfnamefont {D.~C.}\ \bibnamefont
  {Guariento}},\ }\bibinfo {title} {{Hamiltonian analysis of the cuscuton}},\
  \href {https://doi.org/10.1103/PhysRevD.95.104049} {\bibfield  {journal}
  {\bibinfo  {journal} {Phys. Rev. D}\ }\textbf {\bibinfo {volume} {95}},\
  \bibinfo {pages} {104049} (\bibinfo {year} {2017})},\ \Eprint
  {https://arxiv.org/abs/1703.08226} {arXiv:1703.08226 [gr-qc]} \BibitemShut
  {NoStop}%
\bibitem [{\citenamefont {Iyonaga}\ {\it et~al.}(2018)\citenamefont {Iyonaga},
  \citenamefont {Takahashi},\ and\ \citenamefont
  {Kobayashi}}]{Iyonaga:2018vnu}%
  \BibitemOpen
  \bibfield  {author} {\bibinfo {author} {\bibfnamefont {A.}~\bibnamefont
  {Iyonaga}}, \bibinfo {author} {\bibfnamefont {K.}~\bibnamefont {Takahashi}},\
  and\ \bibinfo {author} {\bibfnamefont {T.}~\bibnamefont {Kobayashi}},\
  }\bibinfo {title} {{Extended Cuscuton: Formulation}},\ \href
  {https://doi.org/10.1088/1475-7516/2018/12/002} {J. Cosmol. Astropart. Phys.\
  \bibinfo {volume} {12}\bibfield  {year} {\bibinfo  {year} { (\textbf
  {2018})}\ }\bibfield  {pages} {\bibinfo  {pages} {002}},\ }\Eprint
  {https://arxiv.org/abs/1809.10935} {arXiv:1809.10935 [gr-qc]} \BibitemShut
  {NoStop}%
\bibitem [{\citenamefont {Arnowitt}\ {\it et~al.}(1959)\citenamefont
  {Arnowitt}, \citenamefont {Deser},\ and\ \citenamefont
  {Misner}}]{Arnowitt:1959ah}%
  \BibitemOpen
  \bibfield  {author} {\bibinfo {author} {\bibfnamefont {R.~L.}\ \bibnamefont
  {Arnowitt}}, \bibinfo {author} {\bibfnamefont {S.}~\bibnamefont {Deser}},\
  and\ \bibinfo {author} {\bibfnamefont {C.~W.}\ \bibnamefont {Misner}},\
  }\bibinfo {title} {{Dynamical Structure and Definition of Energy in General
  Relativity}},\ \href {https://doi.org/10.1103/PhysRev.116.1322} {\bibfield
  {journal} {\bibinfo  {journal} {Phys. Rev.}\ }\textbf {\bibinfo {volume}
  {116}},\ \bibinfo {pages} {1322} (\bibinfo {year} {1959})}\BibitemShut
  {NoStop}%
\bibitem [{\citenamefont {Mukherjee}\ and\ \citenamefont
  {Saha}(2009)}]{Mukherjee:2007yi}%
  \BibitemOpen
  \bibfield  {author} {\bibinfo {author} {\bibfnamefont {P.}~\bibnamefont
  {Mukherjee}}\ and\ \bibinfo {author} {\bibfnamefont {A.}~\bibnamefont
  {Saha}},\ }\bibinfo {title} {{Gauge invariances vis-a-vis diffeomorphisms in
  second order metric gravity}},\ \href
  {https://doi.org/10.1142/S0217751X09044759} {\bibfield  {journal} {\bibinfo
  {journal} {Int. J. Mod. Phys. A}\ }\textbf {\bibinfo {volume} {24}},\
  \bibinfo {pages} {4305} (\bibinfo {year} {2009})},\ \Eprint
  {https://arxiv.org/abs/0705.4358} {arXiv:0705.4358 [hep-th]} \BibitemShut
  {NoStop}%
\bibitem [{\citenamefont {Dom\`enech}\ {\it et~al.}(2015)\citenamefont
  {Dom\`enech}, \citenamefont {Mukohyama}, \citenamefont {Namba}, \citenamefont
  {Naruko}, \citenamefont {Saitou},\ and\ \citenamefont
  {Watanabe}}]{Domenech:2015tca}%
  \BibitemOpen
  \bibfield  {author} {\bibinfo {author} {\bibfnamefont {G.}~\bibnamefont
  {Dom\`enech}}, \bibinfo {author} {\bibfnamefont {S.}~\bibnamefont
  {Mukohyama}}, \bibinfo {author} {\bibfnamefont {R.}~\bibnamefont {Namba}},
  \bibinfo {author} {\bibfnamefont {A.}~\bibnamefont {Naruko}}, \bibinfo
  {author} {\bibfnamefont {R.}~\bibnamefont {Saitou}},\ and\ \bibinfo {author}
  {\bibfnamefont {Y.}~\bibnamefont {Watanabe}},\ }\bibinfo {title}
  {{Derivative-dependent metric transformation and physical degrees of
  freedom}},\ \href {https://doi.org/10.1103/PhysRevD.92.084027} {\bibfield
  {journal} {\bibinfo  {journal} {Phys. Rev. D}\ }\textbf {\bibinfo {volume}
  {92}},\ \bibinfo {pages} {084027} (\bibinfo {year} {2015})},\ \Eprint
  {https://arxiv.org/abs/1507.05390} {arXiv:1507.05390 [hep-th]} \BibitemShut
  {NoStop}%
\bibitem [{\citenamefont {Takahashi}\ {\it et~al.}(2017)\citenamefont
  {Takahashi}, \citenamefont {Motohashi}, \citenamefont {Suyama},\ and\
  \citenamefont {Kobayashi}}]{Takahashi:2017zgr}%
  \BibitemOpen
  \bibfield  {author} {\bibinfo {author} {\bibfnamefont {K.}~\bibnamefont
  {Takahashi}}, \bibinfo {author} {\bibfnamefont {H.}~\bibnamefont
  {Motohashi}}, \bibinfo {author} {\bibfnamefont {T.}~\bibnamefont {Suyama}},\
  and\ \bibinfo {author} {\bibfnamefont {T.}~\bibnamefont {Kobayashi}},\
  }\bibinfo {title} {{General invertible transformation and physical degrees of
  freedom}},\ \href {https://doi.org/10.1103/PhysRevD.95.084053} {\bibfield
  {journal} {\bibinfo  {journal} {Phys. Rev. D}\ }\textbf {\bibinfo {volume}
  {95}},\ \bibinfo {pages} {084053} (\bibinfo {year} {2017})},\ \Eprint
  {https://arxiv.org/abs/1702.01849} {arXiv:1702.01849 [gr-qc]} \BibitemShut
  {NoStop}%
\end{thebibliography}

%

\end{document}